\documentclass[sigconf,9pt]{acmart}

\usepackage{subfigure}
\usepackage{pifont}
\usepackage{comment}
\usepackage{color}
\usepackage{booktabs}
\usepackage{balance}
\usepackage{url}
\usepackage{multirow}
\usepackage{amsmath}
\newcommand{\vecg}[1]{\boldsymbol{#1}} 
\renewcommand{\vec}[1]{\mathbf{#1}} 

\newcommand{\blue}{}
\newcommand{\green}{}
\newcommand{\sect}{\textsection}


\copyrightyear{2022}
\acmYear{2022}
\setcopyright{acmcopyright}\acmConference[SenSys '22]{The 20th ACM
	Conference on Embedded Networked Sensor Systems}{November 6--9,
	2022}{Boston, MA, USA}
\acmBooktitle{The 20th ACM Conference on Embedded Networked Sensor
	Systems (SenSys '22), November 6--9, 2022, Boston, MA, USA}
\acmPrice{15.00}
\acmDOI{10.1145/3560905.3568531}
\acmISBN{978-1-4503-9886-2/22/11}

\begin{document}

\title{PriMask: Cascadable and Collusion-Resilient Data Masking for Mobile Cloud Inference\vspace{0em}}

\author{Linshan Jiang, Qun Song, Rui Tan, Mo Li}
\affiliation{\institution{Nanyang Technological University, Singapore}\country{}}




\begin{abstract}
  Mobile cloud offloading is indispensable for inference tasks based on large-scale deep models. However, transmitting privacy-rich inference data to the cloud incurs concerns. This paper presents the design of a system called {\em PriMask}, in which the mobile device uses a secret small-scale neural network called {\em MaskNet} to mask the data before transmission. PriMask significantly weakens the cloud's capability to recover the data or extract certain private attributes. The MaskNet is {\em cascadable} in that the mobile can opt in to or out of its use seamlessly without any modifications to the cloud's inference service. Moreover, the mobiles use different MaskNets, such that the collusion between the cloud and some mobiles does not weaken the protection for other mobiles. We devise a {\em split adversarial learning} method to train a neural network that generates a new MaskNet quickly (within two seconds) at run time. We apply PriMask to three mobile sensing applications with diverse modalities and complexities, i.e., human activity recognition, urban environment crowdsensing, and driver behavior recognition. Results show PriMask's effectiveness in all the three applications.
 \end{abstract}



\maketitle

\section{Introduction}
\label{sec:intro}


Recent years have witnessed the forming fabric of machine learning and mobile computing. While running neural networks on resource-constrained mobile devices has received extensive research \cite{lane2015can,yao2018fastdeepiot}, large-scale neural networks may still incur lengthy execution times that impede user experiences and drain excessive battery energy. {\blue For instance, on Huawei P20 Pro, recognizing a face using Inc-ResNet-V1 with hardware acceleration requires 26 seconds \cite{ignatov2018ai}}.
Such neural networks and those beyond the mobiles' capabilities should run in the cloud. Besides technical constraints, neural networks may have high values and design costs (e.g., 1.3 million US\$ cost for training a natural language processing model \cite{strubell2019energy}).
Thus, many inference services are proprietary and remain in the owners' clouds. As such, {\em cloud inference} is indispensable.

To use cloud inference, the mobile sends the inference sample to the Inference Service Provider (ISP) and receives the result. As mobiles are often used in private spaces and times, the samples may contain privacy. For instance, the voice samples for using a virtual assistant contain rich information about the user, e.g., gender, age, mood, and voiceprint.
Although the network transmissions can be protected by cryptography against external eavesdroppers, protecting the user's privacy against an honest-but-curious ISP while maintaining the accuracy of the cloud inference is a challenging problem. While the ISP honestly executes inference, it may purposely or accidentally extract the users' privacy.


  

To achieve privacy-preserving inference, {\em homomorphic encryption} and {\em neural network masking} approaches have been proposed. In the homomorphic encryption approaches \cite{graepel2012ml,zhan2005privacy,qi2008efficient,gilad2016cryptonets}, the data owner sends the homomorphically encrypted sample to the ISP for performing inference in the encryption domain. However, for resource-constrained mobiles, homomorphic encryption incurs high computation overhead. For instance, it takes more than ten minutes for a $900\,\text{MHz}$ quad-core processor to encrypt a $28\times 28$ grayscale image \cite{jiang2019lightweight}. Differently, neural network masking views several neural network layers as a data masking operation. Thus, the data owner runs these layers and sends the output to the ISP that runs the inference neural network (InferNet) on the masked data.
However, the existing masking approaches \cite{osia2017hybrid,wang2018not,li2021deepobfuscator,liu2019privacy,chi2018privacy,li2020tiprdc} only counteract the privacy threat from the external eavesdroppers who do not have the details of the masking. {\green An eavesdropping attack occurs when an attacker intercepts, deletes, or modifies data that is transmitted between two devices \cite{peake2005eavesdropping,jasim2021eavesdropping,xu2022moving}.  } In above approaches, since the ISP knows the layers used for masking, it can launch the {\em model inversion attack} \cite{he2019model,he2020attacking} to reconstruct the original data.


Running a small-scale neural network has become feasible on mobiles and even lower-profile wearables.
Thus, from the perspective of engineering an workable privacy-respecting cloud inference system for mobiles, neural network masking is a promising basis. In this paper, we design such a system called {\em PriMask} with the following four objectives. {\bf First}, different from the existing studies \cite{osia2017hybrid,wang2018not,li2021deepobfuscator,liu2019privacy,chi2018privacy,li2020tiprdc} that address external eavesdroppers, PriMask considers the privacy threat from the honest-but-curious ISP. {\bf Second}, PriMask is resilient to the collusion between any mobile and ISP, in that the collusion does not weaken the privacy protection for non-colluding mobiles. This collusion-resilient property is important because otherwise the system is susceptible to any compromised individual among many mobiles. {\bf Third}, PriMask does not require any changes to the ISP's InferNet that was designed without privacy preservation considerations. This frees the ISP from the costly redesign and/or retraining of the InferNet. Thus, both the new and legacy mobiles with and without neural network masking, respectively, can coexist in using the cloud inference service. Depending on the remaining battery energy, a mobile can also opt in to or out of PriMask seamlessly without needing to inform the ISP. We say this kind of privacy preservation mechanism {\em cascadable}. {\bf Fourth}, PriMask scales well with the number of mobiles using the inference service.

\begin{figure}
  \subfigure[\small Cascadability ensures no modifications required for the inference neural network (InferNet) running in cloud.]
  {
    \includegraphics[height=.15\textwidth]{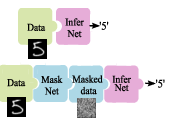}
    \label{fig:cascadable}
  }
  \hfill
  \subfigure[\small Heterogeneity (illustrated by distinct background patterns) provides resilience against the potential collusion.]
  {
    \includegraphics[height=.15\textwidth]{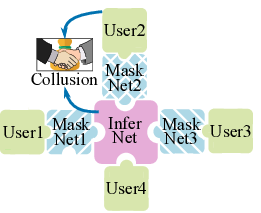}
    \label{fig:anti-collusion}
  }
  \caption{Properties of heterogeneous MaskNets.}
  \vspace{0em}
\end{figure}

We now describe a basic design \cite{xu2020lightweight} to meet the first three objectives except scalability. To implement the cascadable feature, the mobile applies a small-scale mask neural network (MaskNet) on the inference sample and then transmits the output to the ISP. The MaskNet can be obtained by training the concatenation of MaskNet and InferNet. During the training, the InferNet's parameters are fixed and only the training loss is backpropagated from the InferNet to the MaskNet. Fig.~\ref{fig:cascadable} illustrates the cascadable feature achieved by MaskNet. To counteract the privacy threat from the curious ISP, the training is performed by a Privacy Service Provider (PSP) that is trusted by both the mobiles and the ISP. To keep the confidentiality of the ISP's proprietary InferNet, {\em split learning} \cite{vepakomma2018split} is applied such that the MaskNet and InferNet are not revealed to ISP and PSP, respectively. To be collusion-resilient, independent split learning processes can be performed with distinct initialization seeds to yield heterogeneous MaskNets for mobiles.
The heterogeneity is essential to collusion resilience, because otherwise the model inversion attack \cite{he2019model,he2020attacking} launched by the ISP using a colluding mobile's MaskNet is effective to all mobiles using the same MaskNet. Fig.~\ref{fig:anti-collusion} illustrates the heterogeneity that provides the resilience against the potential collusion.

The above basic design requires a separate split training process for each mobile, rendering it non-scalable. For scalability, we advance the design with inspiration from {\em HyperNet} \cite{ha2016hypernetworks}, which is a generative neural network supervising the parameter updates of another neural network.
Given random seeds, a HyperNet generates neural networks with identical architecture but distinct parameters for the same inference task. This is consistent with the heterogeneity requirement for collusion resilience. Thus, if we can replace the process of training a MaskNet in the basic design with inferencing a HyperNet that is much faster, PriMask becomes more scalable in terms of the generation speed of MaskNets. However, we need to address two issues. First, different from the original concept of HyperNet that generates neural networks for a certain inference task, we need a new design of HyperNet to generate MaskNets subject to the ISP's existing InferNet. Second, as HyperNet-generated MaskNets are correlated, the model inversion attack constructed against a specific MaskNet may be transferable to other MaskNets. To address these two issues, we design a {\em split adversarial learning} (SAL) method to train the HyperNet. Specifically, the PSP iteratively trains the HyperNet as defender and an attack neural network (AttackNet) as attacker that implements model inversion or private attribute extraction. During SAL, the HyperNet and InferNet are not revealed to the ISP and PSP, respectively. On the completion of SAL, it is difficult for any adversary who obtains any HyperNet-generated MaskNet to construct an effective AttackNet.

This paper's contributions are summarized as follows:
\begin{itemize}
  \itemsep0em 
\item Different from the existing neural network masking methods \cite{osia2017hybrid,wang2018not,liu2019privacy,li2021deepobfuscator,chi2018privacy,li2020tiprdc} addressing external eavesdroppers, PriMask counteracts curious ISP, mobile-ISP collusion, and requires no changes to the ISP's inference service.
\item We design HyperNet and its SAL training method to improve PriMask's scalability.
  Inferencing the HyperNet to generate a MaskNet takes only milliseconds up to two seconds on a workstation-class PSP.
\item We apply PriMask to three mobile sensing applications with diverse modalities and complexities, i.e., human activity recognition with inertial time series data, urban environment crowdsensing with one-shot tabular data, and driver behavior recognition with image.\footnote{Our implementations are released on the github. https://github.com/jls2007/Primask. } Evaluation shows PriMask's privacy protection performance and scalability to support up to 100,000 mobiles.
\end{itemize}

For simplicity of exposition, this paper assumes that the PSP is trusted by the mobiles. Understood in a different way, PriMask escalates the trustworthiness of the ISP to the level of the PSP. The trustworthiness escalation is useful in practice. For instance, a major cloud computing service provider (e.g., Google) can act as the PSP to escalate the trustworthiness of its small-business tenants who provide inference services. {\green Indeed, Google has provided verification service for small business corporation \cite{google}. Therefore, the potential partners can trust the small business company with verification from Google. Similarly, in our scenario,} the mobile users can trust the inference services at the level that they trust the major cloud computing service provider. {\blue As an analogy in the problem of Internet identity verification, the certificate authorities escalate the trustworthiness of the content providers.} 

Paper organization: \sect\ref{sec:related} reviews related work. \sect\ref{sec:problem} states the problem. \sect\ref{sec:approach} presents the design of PriMask. \sect\ref{sec:case1}, \sect\ref{sec:case2}, and \sect\ref{sec:case3} present the three mobile sensing applications and evaluation. \sect\ref{sec:discuss} discusses related issues. \sect\ref{sec:con} concludes this paper.

\section{Related Work}
\label{sec:related}

This section reviews recent studies on privacy-preserving machine learning and inference.

{\bf Privacy-preserving machine learning} is often based on a system model consisting of \textit{data owners} that contribute training data and a {\em model trainer} that performs or coordinates the model training process.
In the distributed machine learning (DML) schemes including {\em federated learning} \cite{hamm2015crowd,shokri2015privacy,mcmahan2016communication,dean2012large,zinkevich2010parallelized,mo2021ppfl}, the data owners maintain and update local models and exchange the model parameters through the trainer or peer-to-peer communications. As no raw training data are exchanged, DML is considered friendly to privacy-sensitive data owners. To improve DML's privacy preservation, mechanisms including secure aggregation \cite{bonawitz2017practical} and additive perturbation for differential privacy \cite{wei2020federated} have been integrated. To prevent malicious programs from eavesdropping on data in the local training process, the work \cite{mo2021ppfl} applies federated learning in the Trusted Execution Environment (TEE). 
 For mobile devices as data owners, DML imposes high computation overhead due to local model training and also high communication overhead due to its iterative nature. {\blue {\em Split learning} has also been proposed to protect user data privacy \cite{vepakomma2018split,singh2021disco,10.1145/3320269.3384740}. Specifically, it splits a deep neural network into two parts, one at the data owner and the other at the server. Therefore, the server has no direct access to the raw data. However, a recent study \cite{pasquini2021unleashing} finds a privacy vulnerability leakage during split learning. Specifically, it applies adversarial learning to develop a feature-space hijacking attack that shifts the feature domain. Thus, the curious server can reconstruct the training samples or infer their private attributes.  }
 A recent study \cite{rezaei2021application} applies  generative adversarial network (GAN) to address the problem of how a data owner publishes labeled training data with certain private attributes preserved while maintaining the utility of the published data for machine learning. However, as training GAN is highly compute-intensive, this approach is suitable for resource-rich data owners. In addition, the inference model on the published data is jointly trained with the GAN. Thus, the approach is not cascadable.


{\bf Privacy-preserving inference} is based on a similar system model consisting of data owners and an ISP. PriMask belongs to this category.
In CryptoNets \cite{gilad2016cryptonets}, homomorphically encrypted inference sample is transmitted to the ISP for performing inference in the encryption domain.
However, the high computation overhead of homomorphic encryption renders CryptoNets unpractical for resource-constrained devices.
As mentioned in \sect\ref{sec:intro}, neural network masking approaches \cite{osia2017hybrid,wang2018not,li2021deepobfuscator,liu2019privacy,chi2018privacy,li2020tiprdc} have been proposed for privacy-preserving inference.
To enhance preservation strength, the work \cite{osia2017hybrid} uses dimension reduction and Siamese fine-tuning; the work \cite{wang2018not} adopts nullification and additive perturbation for differential privacy.
The studies \cite{liu2019privacy,li2021deepobfuscator,li2020tiprdc}   apply adversarial learning to train the neural network encoder, aiming at negating the adversary's capability of reconstructing the original data or extracting private attributes. However, as discussed in \sect\ref{sec:intro}, these existing approaches \cite{osia2017hybrid,wang2018not,li2021deepobfuscator,liu2019privacy,chi2018privacy,li2020tiprdc} only counteract external eavesdroppers and do not consider the threat from the curious ISP. Because they jointly design the neural network partitions for masking and inference, they are not cascadable. In addition, as they do not impose maskers' heterogeneity, they are susceptible to the collusion between the external eavesdropper and any single data owner. We will show this in \sect\ref{subsec:attacks}.
Besides, they also do not consider fast generation of maskers. Therefore, PriMask differs from these existing approaches \cite{osia2017hybrid,wang2018not,li2021deepobfuscator,liu2019privacy,chi2018privacy,li2020tiprdc} in all four design objectives stated in \sect\ref{sec:intro}. The studies \cite{malekzadeh2019mobile,hajihassnai2021obscurenet} apply autoencoder to publish inertial measurement traces, where the encoder preserves a private attribute and the decoder's output tries to maintain the waveform of the original trace. However, the waveform of many sensing modalities (e.g., image and voice) are privacy-rich, where data publishing is ill-suited.

\section{Problem Statement}
\label{sec:problem}

\subsection{System Overview and Threat Model}
\label{subsec:system model}

\begin{figure}
  \centering
  \includegraphics[width=0.45\textwidth]{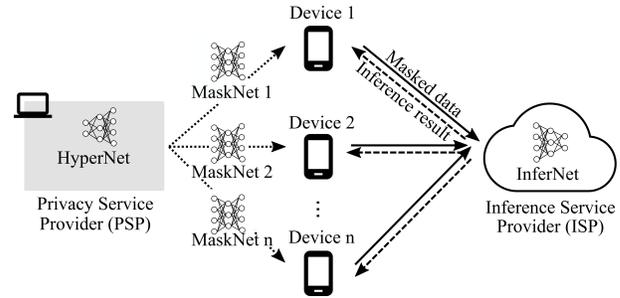}
  \vspace{0em}
  \caption{System model.}
  \label{fig:systemmodel}
  \vspace{0em}
 \end{figure}

As illustrated in Fig.~\ref{fig:systemmodel}, we consider a system consisting of a cloud-based Inference Service Provider (ISP), many mobile devices that desire to use the ISP's service, and a Privacy Service Provider (PSP) that aims at enabling the mobiles to use the ISP's service with certain privacy preserved. The privacy notion will be defined in \sect\ref{subsec:attacks}. We assume that the ISP uses a {\blue pre-trained} deep neural network called InferNet to provide the inference service based on raw input data. InferNet can be large-scale and/or proprietary. Before PSP can serve the mobiles, it works with the ISP by following a protocol to train a neural network called HyperNet. The details of the training approach and the protocol are in \sect\ref{sec:approach}. The mobiles that do not desire privacy preservation can send the raw data to ISP for inferencing InferNet. For each mobile that desires privacy preservation, the PSP inferences the HyperNet with a random seed to generate a small-scale neural network called MaskNet and releases it to the mobile. The MaskNet has identical input and output dimensions. The MaskNets used by the mobiles are identical in architecture but heterogeneous in parameters. When a mobile desires privacy preservation, it feeds the raw data to its MaskNet and sends the output (i.e., masked data) to the ISP. Then, the ISP feeds the masked data to InferNet and returns the result to the mobile. Each mobile should keep its MaskNet confidential.

 PriMask's threat model has three aspects:
 
 $\blacksquare$ {\em Honest-but-curious ISP:} {\green The honest-but-curious (also called semi-honest) adversary is a legitimate participant in a scheme who does not deviate from its function in the defined scheme but attempts to learn extra information from legitimately received messages \cite{mondal2022scotch,malekzadeh2021honest}. In this paper, we consider the honest-but-curious ISP. Specifically,} the ISP honestly executes the InferNet and does not tamper with the received data and the inference results. The ISP also honestly follows the protocol with the PSP to train the HyperNet. However, the ISP is curious about the private information contained in the data received from mobiles and aims at launching privacy attack.
  
 $\blacksquare$ {\em Potential collusion between mobiles and ISP:}  {\green Collusion has been widely studied in the network economics, especially in the auction design \cite{pesendorfer2000study,wu2018cream}. In 2016, a number of water tank suppliers were fined over £2.6 million for breaking competition law by collusion of fixing the price of certain tanks \cite{collusionweb}. In cloud computing and cloud storage, user-server collusion is regarded as a main privacy concern against cloud service providers for searchable symmetric encryption (SSE) \cite{wang2021multi,hamlin2018multi,patel2018symmetric}. In this paper, we consider that} some mobiles may collude with the ISP to find out other mobiles' privacy. Such colluding mobiles surrender their MaskNets to the ISP, which tries to launch privacy attack on non-colluding mobiles. In practice, the ISP may recruit such colluding mobiles by offering monetary benefits. The privacy preservation for such compromised mobiles becomes void. The case in which the ISP pretends mobiles to request MaskNets from the PSP is equivalent to colluding with mobiles. 

 
 $\blacksquare$ {\em PSP trusted by mobiles and ISP:} {\green The trusted third party (TTP) performs important roles in cryptography \cite{zissis2011cryptographic,xu2020trustworthy}. In this paper, we regard the PSP as the TTP. Specifically, }the PSP honestly performs its role, aiming at ensuring the ISP's quality of service while protecting the privacy of the mobiles and ISP. Thus, the PSP is trusted by the mobiles. {\blue Note that, in spite of the trust, PSP shall not request the mobiles' inference data and the ISP's InferNet, which are their role-defining properties in the considered system.} {\green As a service provider, the PSP may gain monetary benefits from the ISP, since privacy-preserving mechanism motivates more users to use the cloud inference service.}

\subsection{Privacy Notation and Attacks}
\label{subsec:attacks}
 PriMask aims at preserving the non-colluding mobiles' privacy from the privacy attacks launched by the ISP solely or in collaboration with the colluding mobiles {\blue during the inference phase}. 
 
{\em Adversary goal:} After receiving the masked inference samples from the non-colluding mobiles, the ISP aims at {\em either} reconstructing the original inference samples, {\em or} extracting a certain private attribute from each masked sample. These two adversary goals are referred to as {\em inversion attack} and {\em private attribute extraction}. {\green Data form confidentiality is an immediate and basic privacy requirement in many applications; private attributes are crucial for mobiles. Note that other privacy attacks (e.g., membership inference attack \cite{nasr2019comprehensive, rigaki2020survey}, model extraction attacks \cite{rigaki2020survey,krishna2019thieves}) are  possible but not the focus of this paper.} The level of privacy protection can be measured by the average dissimilarity between the original and reconstructed samples and the accuracy of the extracted private attributes, respectively. In this paper, we adopt both mean squared error (MSE) and structural similarity index measure (SSIM) as the dissimilarity/similarity metric. Note that recent studies also consider the same privacy notions defined by inversion attack \cite{liu2019privacy,li2021deepobfuscator,he2019model,he2020attacking,chen2020improved,carlini2021private} and private attribute extraction \cite{rezaei2021application,li2021deepobfuscator,hajihassnai2021obscurenet,liu2019privacy, song2019overlearning}.


Now, we discuss the implementations of the privacy attacks after the ISP obtains the MaskNet. The discussions also explain PriMask's system model presented in \sect\ref{subsec:system model}.

\begin{figure}
	\subfigure[\small Original]
	{
		\includegraphics[width=.07\textwidth]{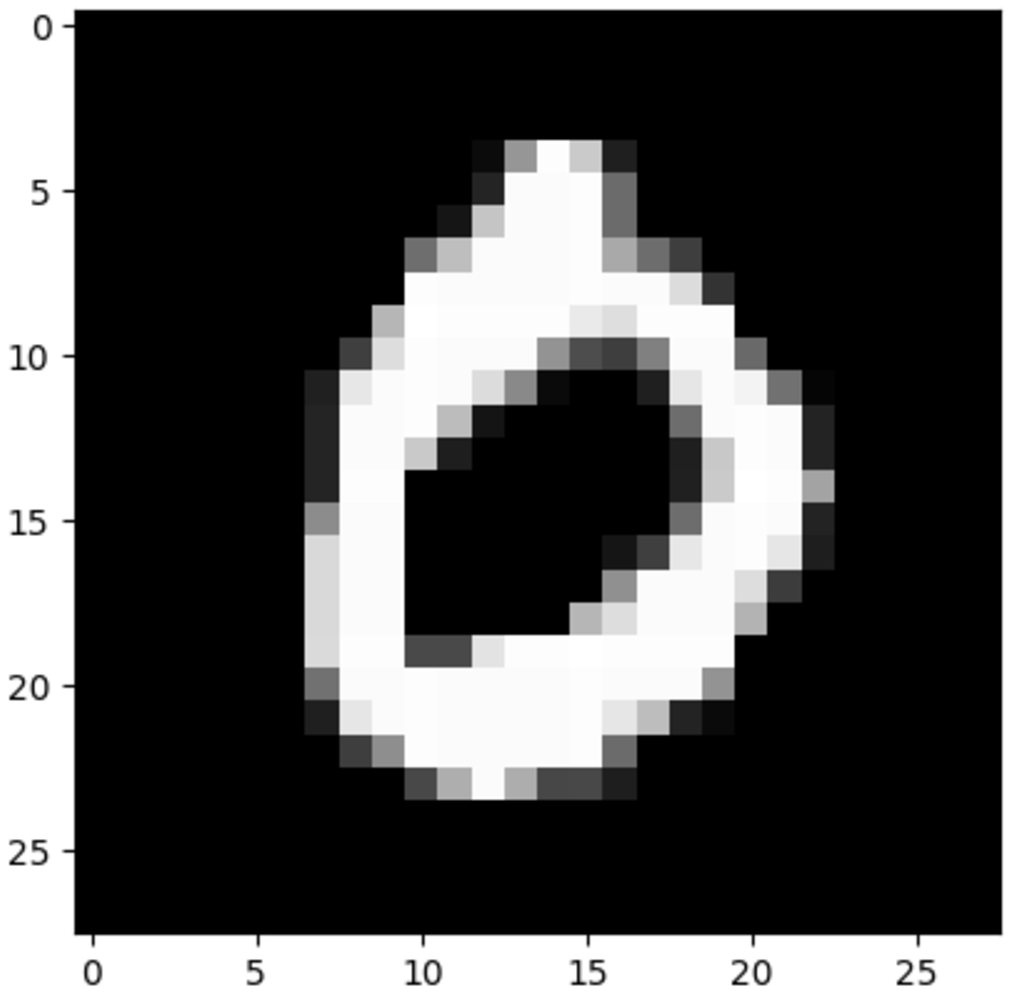}
		\includegraphics[width=.07\textwidth]{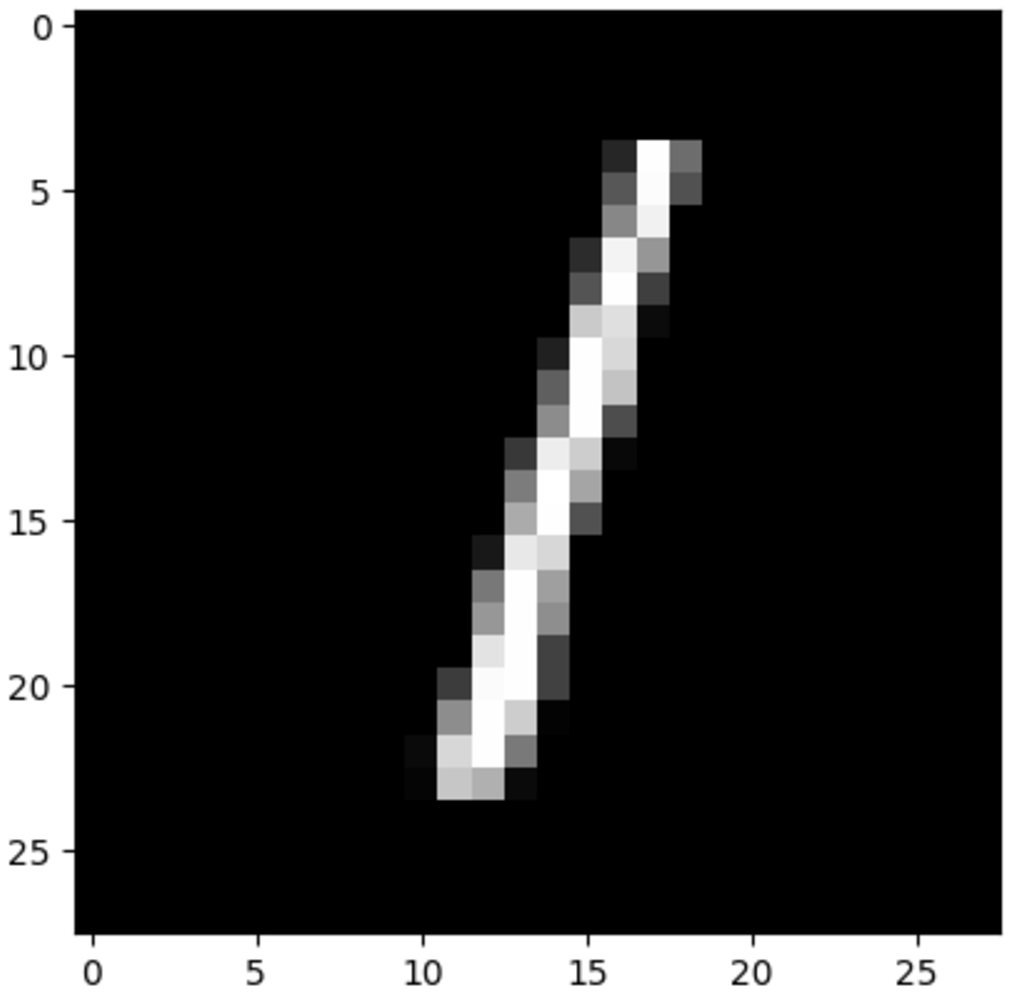}
                \label{fig:mnist-orignial}
	}
	\subfigure[\small Masked]
	{
		\includegraphics[width=.07\textwidth]{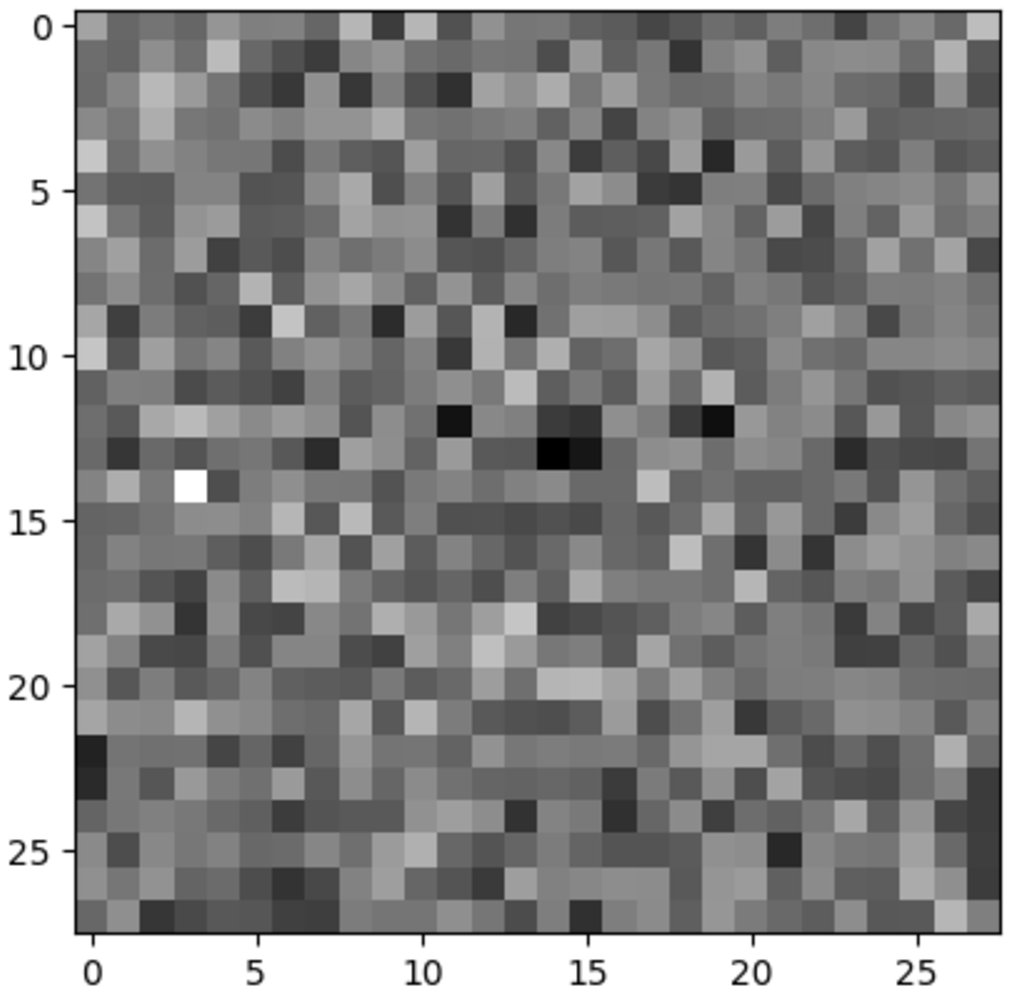}
		\includegraphics[width=.07\textwidth]{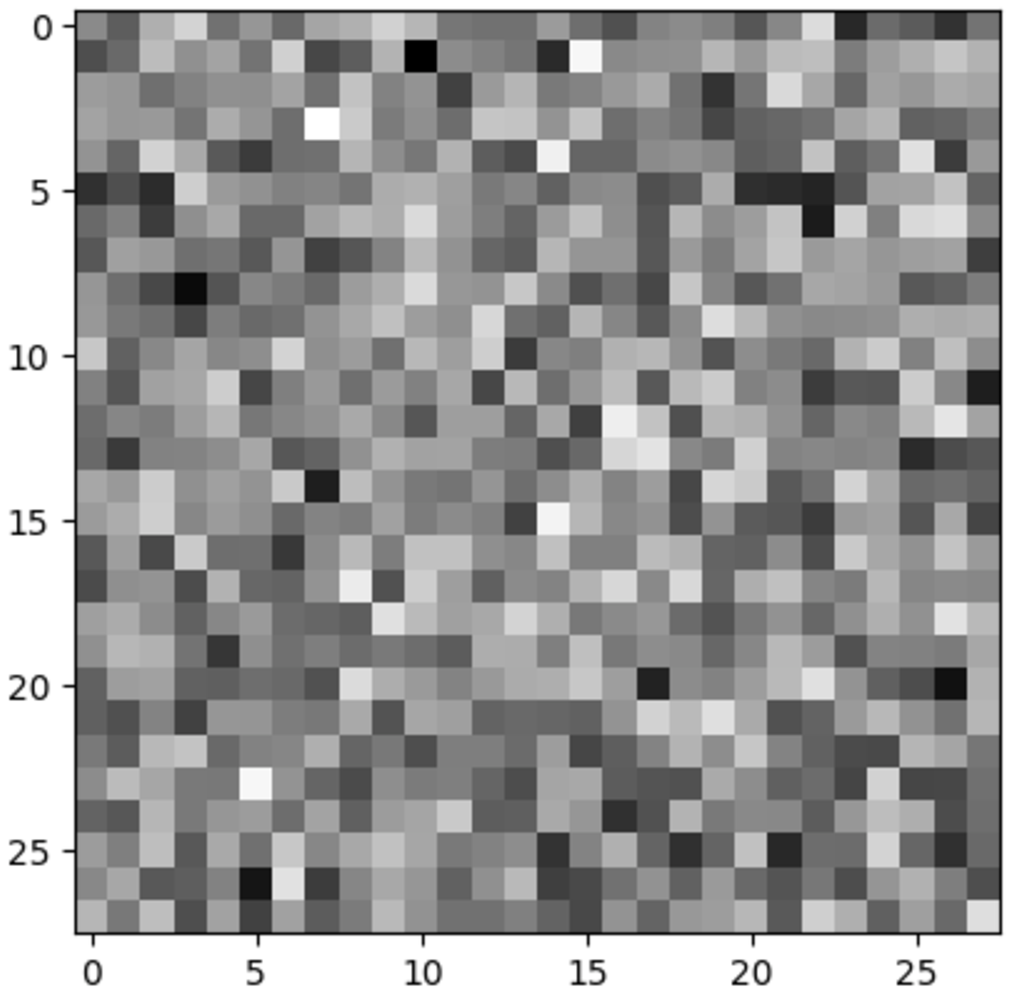}
		\label{fig:mnist-obf}
	}
	\subfigure[\small Reconstructed]
	{
		\includegraphics[width=.07\textwidth]{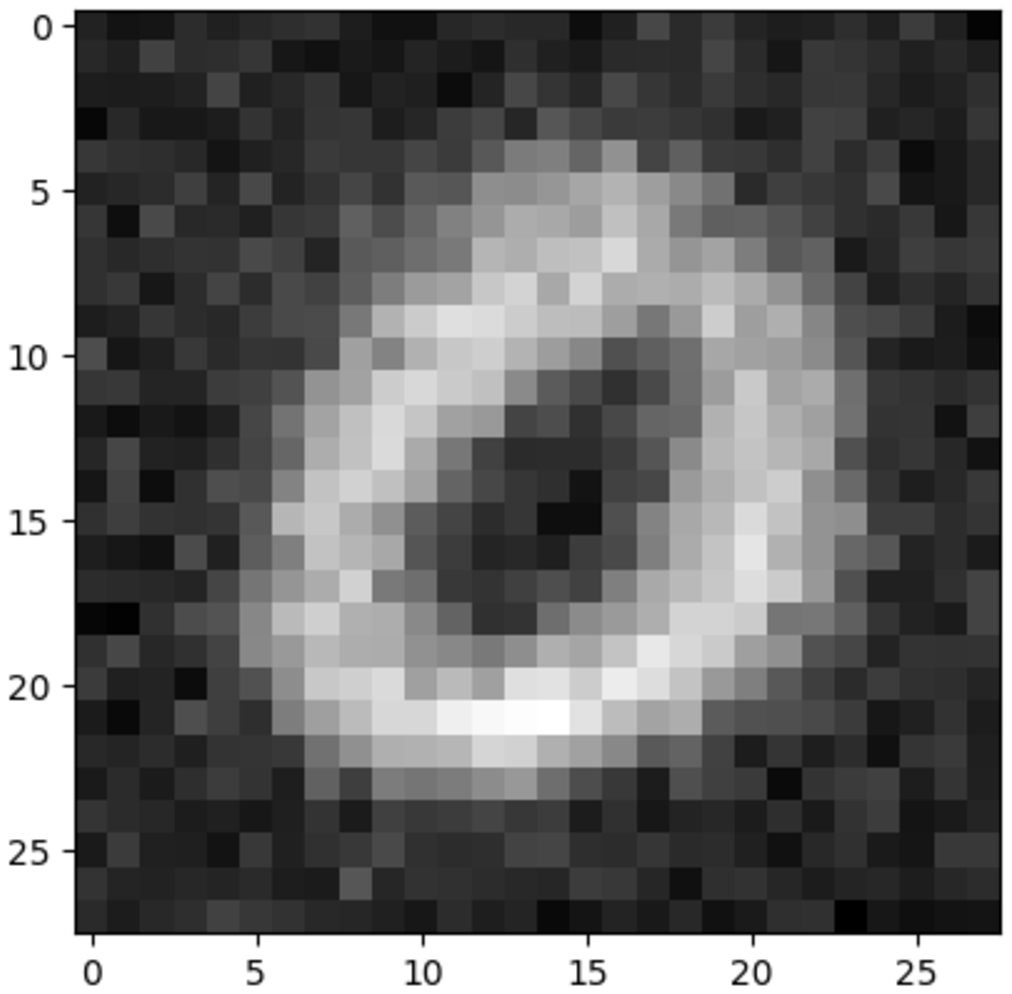}
		\includegraphics[width=.07\textwidth]{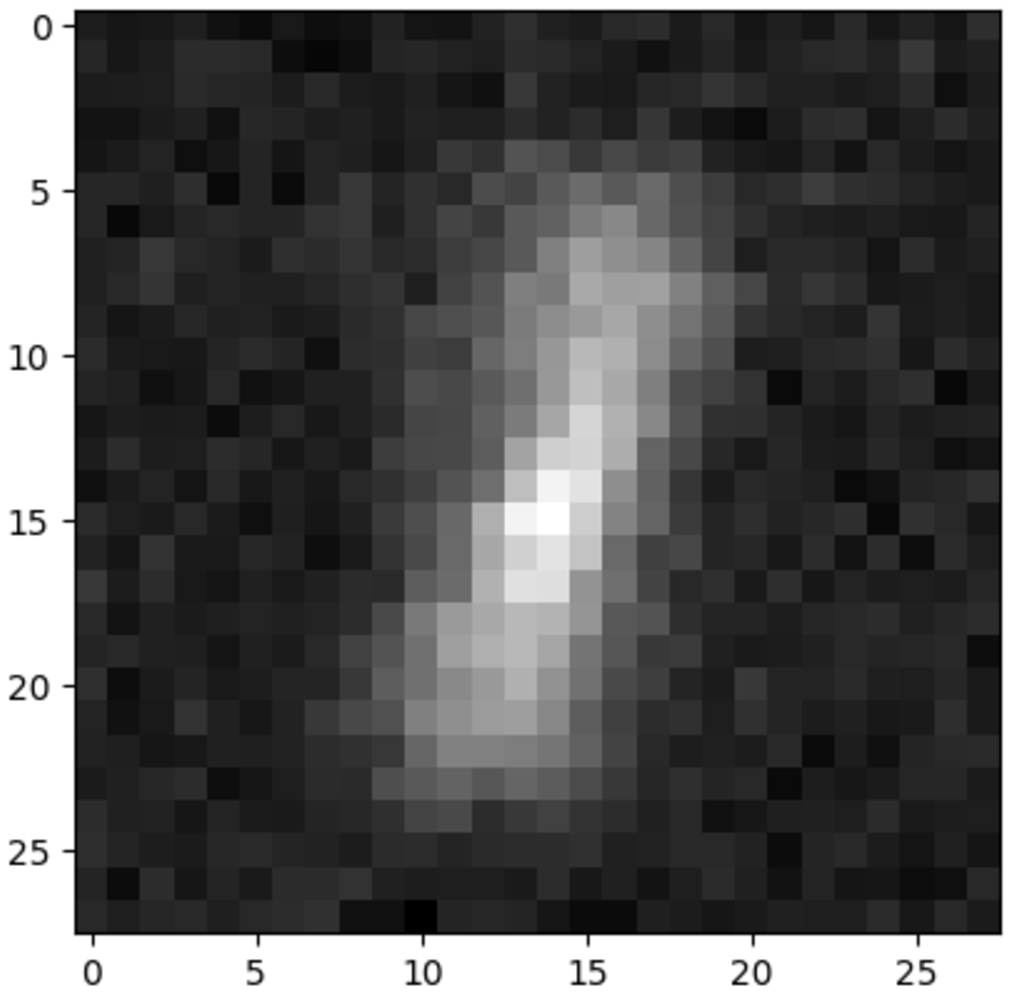}
		\label{fig:mnist-recovered}
              }
              \vspace{0em}
	\caption{Inversion attack on MNIST dataset.}
	\label{fig:mnist-inversion-examples}
        \vspace{0em}
\end{figure}

$\blacksquare$ {\em Inversion attack:} The study \cite{he2019model} presented an inversion attack approach using maximum likelihood estimation. Here we describe a training-based approach. Specifically, the ISP feeds many samples to the MaskNet and obtains the outputs to form a training dataset. Then, the ISP trains a neural network called InvNet that estimates the MaskNet's input from its output. The InvNet can use a mirrored architecture of the MaskNet. Now, we show the effectiveness of the inversion attack using the MNIST handwritten digit dataset \cite{lecun1998mnist}. The MaskNet is generated using our approach described in \sect\ref{sec:approach}. It is a two-layer multilayer perceptron (MLP). The InvNet with a mirrored architecture is trained with MSE of the inversion as the loss function.
Fig.~\ref{fig:mnist-inversion-examples} shows the original, masked, and reconstructed samples for two  digits. The MaskNet can effectively mask the data. However, once the ISP obtains the used MaskNet, it can train the InvNet and reconstruct the original data to certain extents.

$\blacksquare$ {\em Private attribute extraction:} Once the ISP obtains the MaskNet, it can also train a neural network called ExtNet to extract a certain private attribute from the masked data. Specifically, the ISP can feed many samples with private attribute labels to the MaskNet. The MaskNet's outputs labeled with the corresponding private attributes form a training dataset that can be used to train ExtNet. As shown in our human activity recognition application (\sect\ref{sec:case1}), once the ISP obtains the MaskNet, it can train the ExtNet to re-identify the user among 30 users with 63\% accuracy.

\begin{figure*}
  \includegraphics[width=\textwidth]{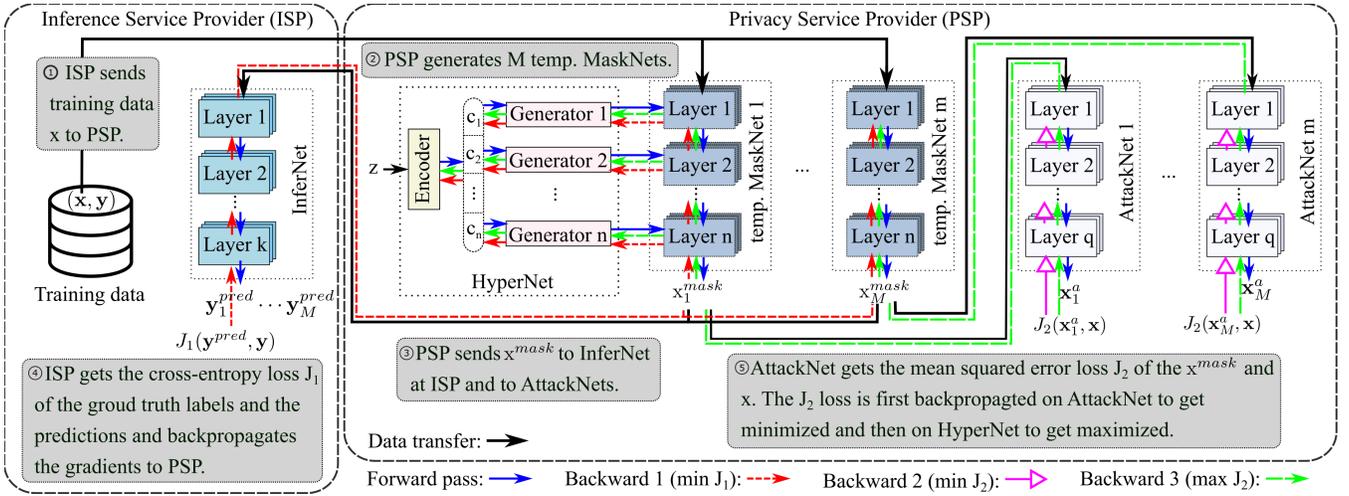}
  \vspace{0em}
  \caption{Proposed split adversarial learning (SAL) framework for training the HyperNet used to generate MaskNets. In SAL, the PSP needs unlabeled training data $\vec{x}$, which can be from an open dataset or the ISP as illustrated in the figure.}
  \label{fig:ourhypernet}
  \vspace{0em}
\end{figure*}

The above privacy attack results give the following implications. First, the generation and release of MaskNets should be performed by the PSP. An authority or a certified organization can be the PSP. Moreover, as discussed in \sect\ref{sec:intro}, a major cloud computing service provider can be the PSP to pass the same trustworthiness on to its small-business tenants that provide inference services.
Second, the mobiles' MaskNets should not be identical. Otherwise, the ISP can launch effective privacy attacks on all mobiles once any one of them colludes with ISP.
Third, the MaskNets and proprietary InferNet should be kept confidential to the ISP and PSP, respectively.

 \subsection{Other Threats}
\label{subsec:other_threat}

{\blue This section discusses other threats that are not addressed by PriMask or irrelevant to PriMask. First, PriMask does not regard ISP's inference result as private information. CryptoNets \cite{gilad2016cryptonets} protects confidentiality of inference result since ISP can only obtain the homomorphically encrypted result. However, the homomorphic encryption of CryptoNets is still not practical for resource-constrained devices. CryptoNets is also not cascadable and requires a redesign of the InferNet. Second, PriMask does not consider the issue of preserving the privacy contained in the training data used by the ISP to pre-train the InferNet, because it is a separate problem and has been studied in literature (e.g., \cite{pasquini2021unleashing,jiang2019lightweight,liu2012cloud,shen2018privacy,wei2020federated,mo2021ppfl}).  Third, the private attribute extraction attack studied in \cite{malekzadeh2021honest} that requires the mobile to execute a maliciously designed deep neural network is irrelevant to PriMask. Specifically, the study \cite{malekzadeh2021honest} proposes an approach to train a privacy-leaking neural network that encodes a private attribute into the neural network's output. The mobile's privacy is compromised if the mobile executes the malicious neural network and transmits the output to a curious party. Differently, in PriMask, the mobile does not execute any maliciously designed neural network. Once the data is masked, the concerned private attribute is preserved. Thus, the attack method in \cite{malekzadeh2021honest} is irrelevant to this paper.}

\section{PriMask Design \& Implementation}
\label{sec:approach}

This section is organized as follows. \sect\ref{subsec:pre} presents the preliminary on HyperNet. \sect\ref{susection:designofheter} presents the split adversarial learning (SAL) framework for the HyperNet used to generate MaskNets. \sect\ref{sucsec:hyper} presents the SAL protocol between the PSP and ISP. \sect\ref{subsec:application-considerations} discusses the generalizability and implementations of PriMask. \sect\ref{subsec:mnist} presents the results on MNIST as a simple case study.

\subsection{Preliminary on HyperNet}
\label{subsec:pre}


HyperNet \cite{ha2016hypernetworks} is a neural network generating the parameters of the target neural network. In this paper, we use HyperNet to generate MaskNets. As illustrated in Fig.~\ref{fig:ourhypernet}, the HyperNet consists of an {\em encoder} network and $n$ {\em weight generator} networks, where the HyperNet's parameters $\vecg{\phi} = \{\vecg{\phi}_E, \vecg{\phi}_G\}$, $\vecg{\phi}_E$ denotes the encoder's parameters, $\vecg{\phi}_G$ denotes all generators' parameters.
The encoder takes as input a random vector $\vec{z}$ sampled from a normal distribution $\mathcal{N}(\vec{0},\vec{I})$. The encoder maps $\vec{z}$ to $n$ latent codes denoted by $\{c_i | i = 1, 2, \ldots, n\}$, which are then fed to $n$ weight generators, respectively. Each weight generator outputs the parameters of a layer of the target neural network.
The above process of inferencing the HyperNet, which is represented by $h(\vec{z};\vecg{\phi})$, can complete in a short time.
By repeating the process with different inputs $\vec{z}$, many distinct neural networks can be generated. {\blue  Such HyperNet-generated neural networks have been applied to detect out-of-distribution inputs \cite{henning2018approximating} and adversarial examples  \cite{ratzlaff2019hypergan}, and assess the uncertainty on an inference sample \cite{wang2018adversarial,su2020blindly}.} The training of HyperNet is addressed in the following subsections.


\subsection{Split Adversarial Learning Framework}
\label{susection:designofheter}

Fig.~\ref{fig:ourhypernet} illustrates the proposed SAL framework for training HyperNet to generate MaskNets. It integrates the principles of split learning \cite{vepakomma2018split} and adversarial learning \cite{huang2011adversarial}. It consists of four modules: InferNet at ISP, HyperNet and $M$ AttackNets at PSP, and $M$ temporary MaskNets generated by HyperNet. The $M$ is a training hyperparameter {\blue that can be dynamically adjusted during training to accelerate convergence. For simplicity, we initially set $M$ equal to the batch size.} The AttackNets trained by the PSP form the adversary of the adversarial learning \cite{huang2011adversarial}, which assists the PSP to train the HyperNet that can generate MaskNets more robust against the privacy attack launched by the ISP. Depending on the privacy protection goal (inversion attack or private attribute extraction), the AttackNet can be either InvNet or ExtNet. The core of SAL is the definitions of the training loss functions.

 
The notation used in this subsection is defined as follows. Denote by $\vec{x} = \{x_1,x_2, \ldots, x_N\}$ the set of training samples, by $\vec{y} = \{y_1,y_2, \ldots,y_N\}$ the corresponding class labels, and by $\vec{a} = \{a_1,a_2,\ldots,a_N\}$ the corresponding private attribute labels, where $N$ is the cardinality of the training dataset. Denote by $f_{\mathrm{Mask}}(\cdot; \vecg{\theta}_m)$ the $m^{\text{th}}$ temporary MaskNet, where $\vecg{\theta}_{m}$ represents the parameters generated by the HyperNet, i.e., $\vecg{\theta}_m = h(\vec{z}_m; \vecg{\phi})$.
Denote by $f_{\mathrm{Inf}}(\cdot;\vecg{\psi})$ the pre-trained proprietary InferNet, where $\vecg{\psi}$ represents the parameters that are constant during SAL.
Denote by $f_{\mathrm{Att}}(\cdot;\vecg{\xi}_m)$ the $m^\text{th}$ AttackNet, where $\vecg{\xi}_m$ represents the parameters. During adversarial learning, the $m^\text{th}$ AttackNet is used as the adversary against the $m^\text{th}$ temporary MaskNet. Denote by $J(\vec{y}^{\mathrm{pred}},\vec{y})$ the cross-entropy loss function, where $\vec{y}$ and $\vec{y}^{\mathrm{pred}}$ are the ground-truth and predicted labels. The cross-entropy loss function also admits privacy labels, i.e., $J(\vec{a}^{\mathrm{pred}},\vec{a})$. Denote by $E(\vec{x}^a, \vec{x})$ the MSE loss, where $\vec{x}$ and $\vec{x}^a$ are the original samples and those reconstructed by InvNet.

Now, we define the loss functions. When a batch of $M$ HyperNet-generated MaskNets are used, the ISP's quality of service is characterized by the following cross-entropy loss:
 \begin{equation}
J_1 = \frac{1}{M}\sum^M_{m=1} J \left( f_{\mathrm{Inf}} \left( f_{\mathrm{Mask}} \left( \vec{x}; h \left( \vec{z}_m;\vecg{\phi} \right) \right); \vecg{\psi} \right), \vec{y} \right).
\label{eq:Infloss}
\end{equation}
Depending on the privacy protection goal, the effectiveness of PSP's $m^\text{th}$ AttackNet is characterized by the following loss:
\begin{equation*}
  J_{2,m}\!( \! \vecg{\xi}_m \! ) \!\! = \!\! \left\{ 
  \begin{array}{ll}
    \!\!\! E  \left( f_{\mathrm{Att}}  \left( f_{\mathrm{Mask}} \left(\vec{x}; h \left(\vec{z}_m;\vecg{\phi} \right) \right) ;\vecg{\xi}_m  \right) , \vec{x} \right), & \!\!\!\!\text{for InvNet;} \\
    \!\!\! J  \left( f_{\mathrm{Att}}  \left (f_{\mathrm{Mask}}  \left( \vec{x}; h \left( \vec{z}_m; \vecg{\phi} \right) \right) ; \vecg{\xi}_m  \right) , \vec{a} \right), & \!\!\!\!\text{for ExtNet.} \\
  \end{array}
  \right. \!
\end{equation*}

The SAL workflow is as follows.
First, HyperNet is trained to minimize $J_1$, i.e.,
$\vecg{\phi}^* = \operatorname*{arg\,min}_{\vecg{\phi}} \mathbb{E}_{\vec{z}_m \sim \mathcal{N}(\vec{0}, \vec{I}), \forall m \in [1,M]} \left[ J_1 \right]$.
Then, each AttackNet is trained to minimize $J_{2,m}$ against the corresponding temporary HyperNet-generated MaskNet:
$\vecg{\xi}_{m}^* =  \operatorname*{arg\,min}_{\vecg{\xi}_{m}} J_{2,m} (\vecg{\xi}_{m})$, $\forall m \in [1, M]$.
Lastly, the HyperNet is refined to achieve a multi-objective goal of minimizing $J_1$ and maximizing each $J_{2,m}(\vecg{\xi}_m^{*})$, where the latter aims at defeating the privacy attack. We represent the multi-objective goal using a single composite loss function to direct the refinement of the HyperNet:
\begin{equation}
\vecg{\phi}^{**} \!=\!  \operatorname*{arg\,min}_{\vecg{\phi}} \mathbb{E}_{\vec{z}_m \sim \mathcal{N}(\vec{0},\vec{I})}\left[ J_1 \!-\! \frac{\lambda}{M} \!\sum_{m=1}^{M} J_{2,m}( \vecg{\xi}_{m}^* ) \right],
\label{eq:mixedlossrec}
\end{equation}
where the {\em adversarial learning factor} $\lambda$ balances the objectives of maintaining the ISP's quality of service and defeating the privacy attack. The $\lambda$ can be used to tune the trade-off between the inference service quality and the privacy protection level. In \sect\ref{sucsec:hyper}, we will discuss how to set $\lambda$.

{\blue The existing studies \cite{henning2018approximating,tang2021learning} also apply adversarial learning to train HyperNets, for approximating predictive distributions \cite{henning2018approximating} or capturing complex policy distributions \cite{tang2021learning}. In these studies, a single distribution discriminator is used as the adversary of the adversarial learning. Differently, in PriMask's SAL, each temporary MaskNet has a separate AttackNet, which is consistent with the fact that the ISP can craft custom AttackNet for any obtained MaskNet.}

\subsection{Split Adversarial Learning Protocol}
\label{sucsec:hyper}


This section presents the protocol between ISP and PSP to implement SAL. To drive SAL, the PSP needs to feed unlabeled data samples $\vec{X}$ from the training dataset $(\vec{X}, \vec{Y})$ to the HyperNet-generated MaskNets. If the training dataset is not publicly available, the ISP transmits $\vec{X}$ (excluding $\vec{Y}$) to the PSP, as illustrated by step \ding{192} in Fig.~\ref{fig:ourhypernet}.
Then, ISP and PSP start training.
In each loop of a training epoch, the SAL protocol has the following three phases.

{\bf (1) Updating HyperNet:}
The PSP samples a mini-batch $\vec{x}$ from $\vec{X}$. On $\vec{x}$, the PSP draws $M$ random vectors $\vec{z}_1$, $\vec{z}_2$, $\ldots$, $\vec{z}_M$ from $\mathcal{N}(\vec{0}, \vec{I})$ and generates $M$ MaskNets using the current HyperNet, as illustrated by step \ding{193} in Fig.~\ref{fig:ourhypernet}. The masked mini-batch by the $m^\text{th}$ MaskNet is denoted by $\vec{x}_{m}^{\mathrm{mask}}$. To compute the gradient of Eq.~(\ref{eq:Infloss}), the PSP sends $\{\vec{x}_{1}^{\mathrm{mask}},\vec{x}_{2}^{\mathrm{mask}}, \ldots,\vec{x}_{m}^{\mathrm{mask}}\}$ to the ISP, as illustrated by step \ding{194} in Fig.~\ref{fig:ourhypernet}. The ISP feeds the received masked data to the InferNet to obtain the predictions $\{\vec{y}_1^{\mathrm{pred}}, \vec{y}_2^{\mathrm{pred}}, \ldots, \vec{y}_M^{\mathrm{pred}}\}$. Then, the ISP computes the cross-entropy loss $J_1$ for the mini-batch using the ground truth labels $\vec{y}$ and sends the backward gradients of $\{\vec{x}_{1}^{\mathrm{mask}},\vec{x}_{2}^{\mathrm{mask}},\ldots,\vec{x}_{m}^{\mathrm{mask}}\}$ to the PSP, as illustrated by step \ding{195} in Fig.~\ref{fig:ourhypernet}. The InferNet's parameters remain unchanged during SAL; they are solely used to compute the backward gradients. Upon receiving the backward gradients, the PSP backpropagates them through the MaskNets without updating their parameters and then through the HyperNet to update its parameters $\vecg{\phi}$ for minimizing the cross-entropy loss $J_1$.

{\bf (2) Updating AttackNets:}
After updating $\vecg{\phi}$ on $\vec{x}$, PSP enters the adversarial learning phase to update the $M$ AttackNets. Specifically, the PSP regenerates $M$ MaskNets using newly sampled random vectors $\{\vec{z}_1,\vec{z}_2,\ldots,\vec{z}_M\}$ and the latest $\vecg{\phi}$.
The masked mini-batch $\vec{x}_{m}^{\mathrm{mask}}$ produced by the $m^\text{th}$ updated MaskNet is fed into the corresponding AttackNet, as illustrated by step \ding{194} in Fig.~\ref{fig:ourhypernet}. The MSE/cross-entropy loss is backpropagated to update the AttackNets' parameters $\{\vecg{\xi}_1, \vecg{\xi}_2, \ldots, \vecg{\xi}_M\}$ to minimize $J_{2,m}$,
as illustrated by step \ding{196} in Fig.~\ref{fig:ourhypernet}. For the same mini-match, the PSP repeats the above process for multiple times to gain better AttackNets.


{\bf (3) Refining HyperNet:}
The last phase on the current mini-batch $\vec{x}$ is to refine the HyperNet according to the composite loss function in Eq.~(\ref{eq:mixedlossrec}). Similar to Phase (1), the PSP sends the latest masked data samples $\{\vec{x}_{1}^{\mathrm{mask}},\vec{x}_{2}^{\mathrm{mask}}, \ldots,\vec{x}_{m}^{\mathrm{mask}}\}$ to the ISP and receives the backward gradients corresponding to the loss $J_1$. The PSP also computes the backward gradients of the AttackNets corresponding to the losses $\{J_{2,1}, J_{2,2},\ldots,J_{2,M}\}$. PSP updates HyperNet's parameters $\vecg{\phi}$ according to Eq.~(\ref{eq:mixedlossrec}).

The PSP repeats the above three phases on multiple mini-batches in the current training epoch. Once all the training samples are utilized in the current epoch, the PSP proceeds to the next epoch. Upon the completion of the training, the PSP is ready to serve the mobiles. Specifically, to respond to a mobile's service request, the PSP feeds a random vector $\vec{z}$ sampled from $\mathcal{N}(\vec{0}, \vec{I})$ to the HyperNet to generate a MaskNet and releases it to the mobile.

Now, we discuss the setting of the adversarial learning factor $\lambda$. {\blue The setting of this factor $\lambda$ is task-specific and may not transfer across inference tasks.} The PSP may perform SAL for multiple rounds with different $\lambda$ settings to train multiple HyperNets. For each SAL process, the PSP may measure the average test accuracy and the metric characterizing the effectiveness of the privacy attack (e.g., the average MSE of the inversion attack on non-colluding mobiles). {\green Based on the validation on  average test accuracy and estimation of the privacy attack,} the PSP can publish a table of suitable $\lambda$ settings and the associated test accuracy and privacy attack effectiveness metric. Each mobile may inform the PSP with its preferred $\lambda$ setting and obtain a MaskNet generated by the PSP using the corresponding HyperNet. {\green We generate the table of $\lambda$ settings and corresponding metrics on MNIST in \sect\ref{subsec:mnist}.}

\subsection{Generalizability and Implementations}
\label{subsec:application-considerations}

  \begin{table}[ht]
    \caption{PriMask applications and benchmark results including model sizes, training time (unit: minute), bandwidth usage (unit: Mb/s), communication volumes per sub-epoch and compute times (unit: millisecond).}
    \vspace{0em}
    \label{tab:compute-times}
    \small
    \begin{tabular}{lcccc}
      \toprule
      {\bf Application} & {\bf MNIST} & {\bf HAR} & {\bf UEC} & {\bf DBR} \\      
      \midrule
      Sensor & camera & IMU & multiple$^*$$\!\!\!\!\!$ & camera \\
      Data type & 28x28 & time & one-shot & 240x240$\!\!\!\!\!$\\
      & image & series & tabular & image \\
      Sample size & 784 & 1,152 & 5 & 57,600 \\
      Private attribute & n/a & identity & location & identity \\
      \midrule
      InferNet size (MB) & 0.08 & 8.35 & 0.06 & 226.37 \\
      HyperNet size (MB)$\!\!\!\!\!\!\!$ & 12.64 & 113.01 & 0.20 & 3310 \\
      MaskNet size (MB) & 0.20 & 1.76 & 0.003 & 54 \\
      \midrule
      HyperNet training$^\dag$ $\!\!\!\!\!\!\!$ & 70 & 95 & 10& $\sim$7200 \\
      Comm. vol. (MB) & 7.58 & 11.13 & 0.04 & 556.89 \\
      Bandwidth Usage $\!\!\!\!\!\!\!$ & 3.375 & 0.69 & 0.04& 0.875 \\
      \midrule
      MaskNet generation$^\dag$ & 1.5 & 2.3 & 1.6 & 2130 \\
      MaskNet execution$^\ddag$ & 0.011 & 1.33 & 0.03 & 42.02 \\
      MaskNet execution$^\star$ & 0.078 & 1.14 & 0.05 & 21.24 \\
      \bottomrule
      \multicolumn{5}{l}{$^*$Light, microphone, air pressure, temperature.}\\
      \multicolumn{5}{l}{$^\dag$On two computers with i7-6850K CPUs and  Quadro RTX 6000 GPUs.}\\
      \multicolumn{5}{l}{$^\ddag$On Jetson Nano's quad-core Cortex-A57 processor.}\\
      \multicolumn{5}{l}{$^\star$On Google Pixels 4's octa-core Qualcomm Snapdragon 855 processor.}
    \end{tabular}
    \vspace{0em}
  \end{table}
  
  As SAL is agnostic to InferNet, PriMask can be applied to different inference tasks. In \sect\ref{subsec:mnist}, we apply PriMask to a simple handwritten digit recognition task as a starting case study. In \sect\ref{sec:case1}, \sect\ref{sec:case2}, and \sect\ref{sec:case3}, we apply PriMask to three mobile sensing tasks of human activity recognition (HAR), urban environment crowdsensing (UEC), and driver behavior recognition (DBR). As summarized in Table~\ref{tab:compute-times}, the three applications have diverse sensing modalities, data types, and InferNet complexities. All these four applications use similar MaskNet and HyperNet architectures. Therefore, PriMask has good generalizability.

  We use PyTorch \cite{pytorch} to implement the InferNets and HyperNets on workstation computers. We use PyTorch and PyTorch Mobile \cite{pytorch-mobile} to implement the {\blue on-device} MaskNet-based data masking on Jetson Nano \cite{nano} running Ubuntu OS and Google Pixel4 smartphone running Android OS, respectively.
  Table~\ref{tab:compute-times} shows the sizes of the models used by PriMask and the compute overhead measurements for all the four applications. {\blue If no privacy protection is implemented, the mobile transmits raw inference data to the ISP. Thus, the computation overhead in Table~\ref{tab:compute-times} fully attributes to PriMask. }{\blue The HyperNet training time is the average value over 10 rounds;} the MaskNet generation and execution times are average values over 1,000 executions.  For each application, the size of HyperNet is larger than the size of InferNet. As HyperNets are executed on PSP's server-class computers, their large sizes are not a concern. The time for executing HyperNet to generate a MaskNet is at most 2.13 seconds. Executing MaskNet to mask a sample on Jetson Nano and Pixel4 just takes tens of milliseconds at most, representing low overheads. {\blue Note that the training of HyperNet involves communications between PSP and ISP. The maximum bandwidth usage of $3.375\,\text{Mb/s}$ is not a high overhead, especially in today's Wi-Fi/5G environments.} 



\subsection{A Simple Case Study on MNIST}
\label{subsec:mnist}

\begin{figure}
  \centering
  \includegraphics[width=0.4\textwidth]{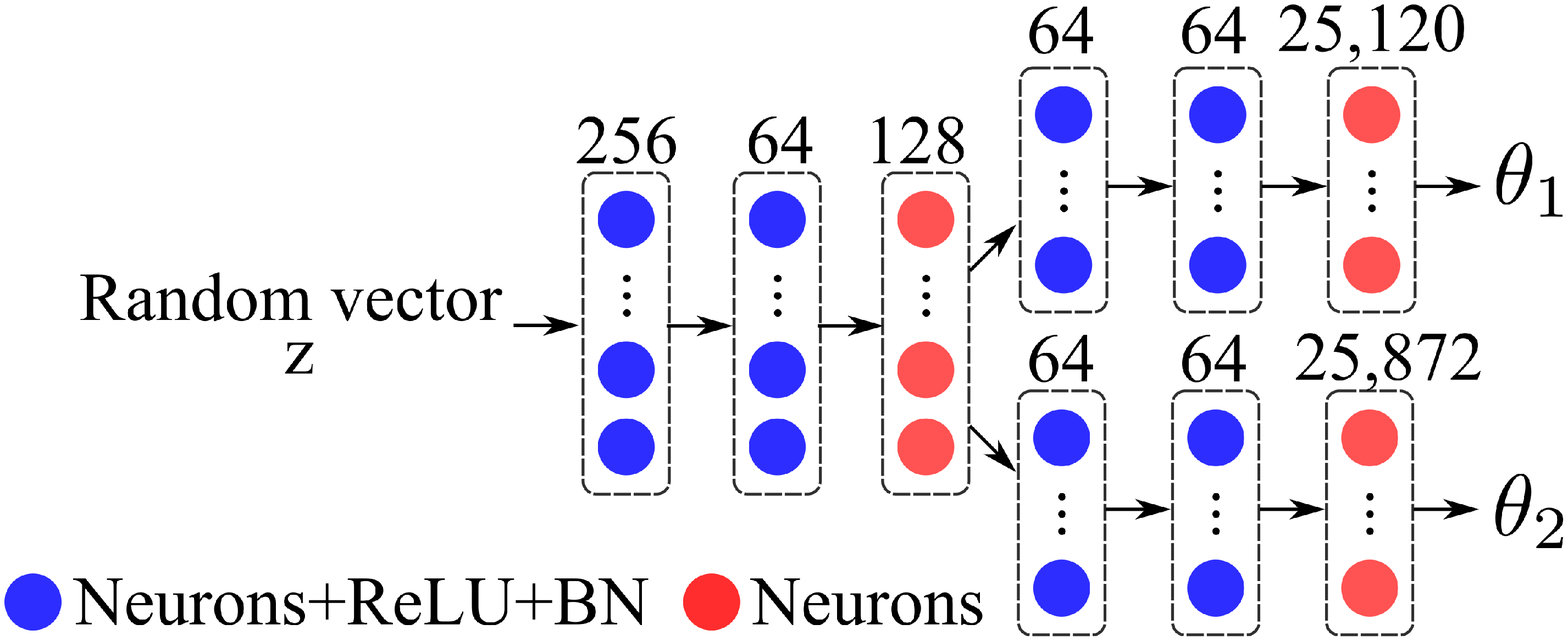}
  \vspace{0em}	
  \caption{HyperNet architecture.}
  \label{fig:mnisthyper}
  \vspace{0em}
\end{figure}

As the MNIST dataset facilitates visualization, we use it as a starting and simple case study.
MNIST consists of 60,000 training samples and 10,000 testing samples.
Each sample is a $28 \times 28$ image showing a handwritten digit.
The pixel value is normalized to $[0, 1]$.
We design a convolutional neural network (CNN) as the InferNet.
The CNN consists of two convolutional layers with max pooling, three dense layers with Rectified Linear Unit (ReLU) activation, and a softmax function to generate the classification result.
The test accuracy of the InferNet on raw testing samples is 98.7\%.
The MaskNet adopts an MLP architecture with a single hidden layer consisting of 32 neurons.
There are 25,120 trainable parameters
between the input and hidden layers, and 25,872 trainable parameters
between the hidden and output layers. Thus, HyperNet's output consists of two parts corresponding to the above two groups of parameters.
Fig.~\ref{fig:mnisthyper} shows the HyperNet's architecture.

\begin{figure}
  \subfigure[CDF of accuracy]{
    \includegraphics[width=0.145\textwidth]{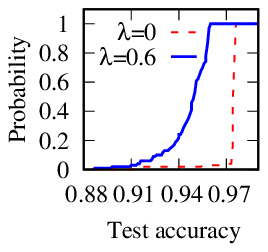}
    \label{fig:testacccdf_mnist}
  }
  \subfigure[CDF of MSE]{
    \includegraphics[width=0.145\textwidth]{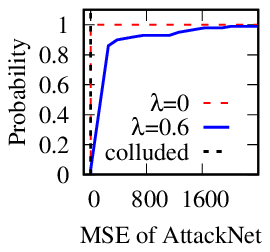}
    \label{fig:reconstructioncdf_mnist}
  }
  \subfigure[CDF of SSIM]{
    \includegraphics[width=0.145\textwidth]{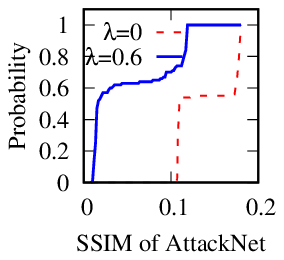}
    \label{fig:reconstructioncdf_mnist_ssim}
  }
  \vspace{0em}
  \caption{Impact of PriMask on MNIST test accuracy and privacy protection. $\lambda$ is adversarial learning factor ($\lambda=0$ means that adversarial learning is not enabled).}
  \label{fig:cdfs_mnist}
  \vspace{0em}
\end{figure}

The training samples including labels are used by ISP to build InferNet. The training samples excluding labels are used by PSP to build HyperNet in collaboration with ISP according to the SAL protocol. To evaluate a MaskNet generated by HyperNet, we feed all test samples to the MaskNet-InferNet pipeline and measure the test accuracy. Since our focus is to understand the impact of MaskNet on the inference, there is no need to simulate the system by assigning disjoint portions of the test dataset to all mobiles. We follow this evaluation methodology throughout this paper.

{\em Impact of PriMask on InferNet accuracy:} We use the HyperNet to generate MaskNets for 100 mobiles. First, we evaluate the InferNet's test accuracies when the 100 MaskNets are used. Fig.~\ref{fig:testacccdf_mnist} shows the cumulative distribution functions (CDFs) of test accuracies when the adversarial learning factor $\lambda=0$ and $\lambda=0.6$. When $\lambda=0$, the adversarial learning of SAL is not enabled. In this case, the InferNet's test accuracies corresponding to the 100 MaskNets are mostly within $(95.5\%, 97.6\%)$, with an average value of 97.2\%. When $\lambda=0.6$, the test accuracies are mostly within $(91.5\%, 95.9\%)$, with an average value of 94.5\%. Compared with the original test accuracy of 98.7\%, PriMask results in average test accuracy losses of 1.5\% and 4.2\%, when the adversarial learning is disabled and enabled, respectively. We will show shortly that the adversarial learning enhances the privacy protection. Therefore, there is a trade-off between maintaining test accuracy and preserving privacy.
\begin{figure}
	\subfigure[\small Original]
	{
          \includegraphics[width=.08\textwidth]{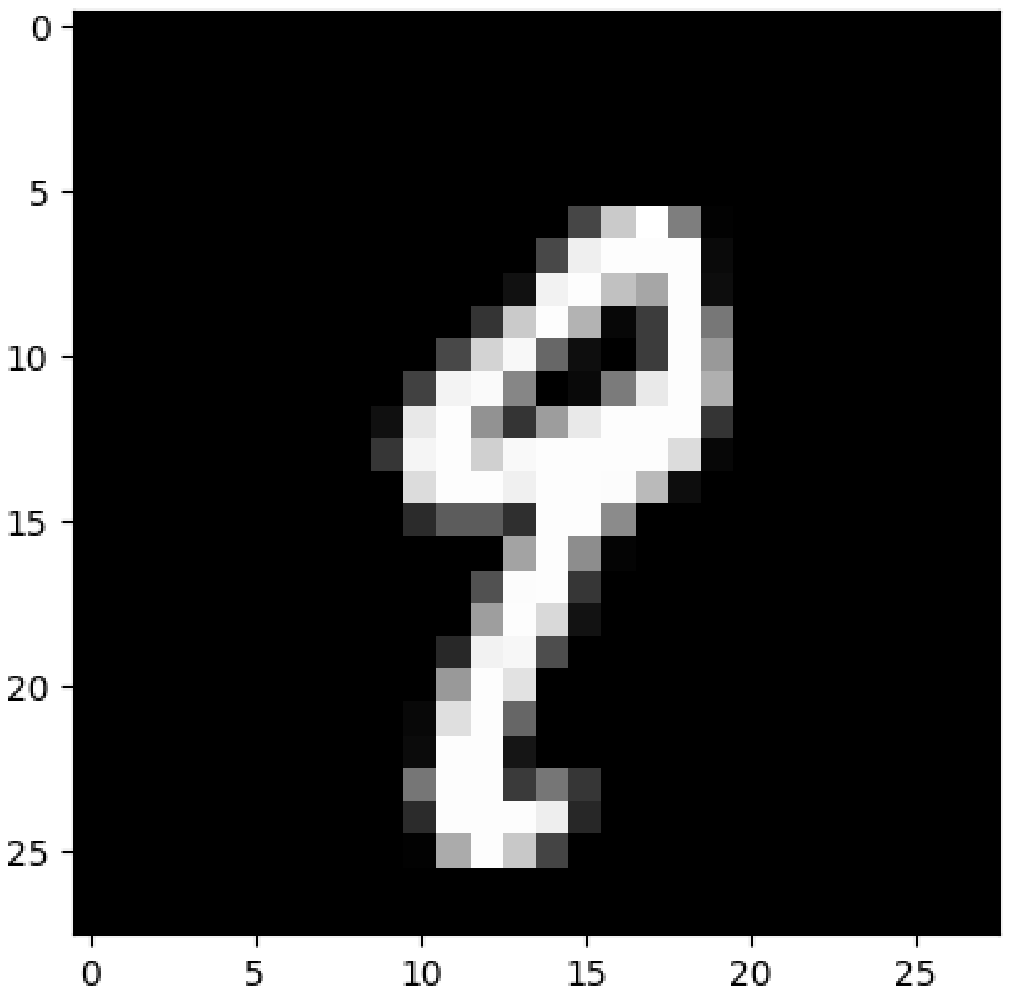}
	}
        \subfigure[\small Masked]
	{
          \includegraphics[width=.08\textwidth]{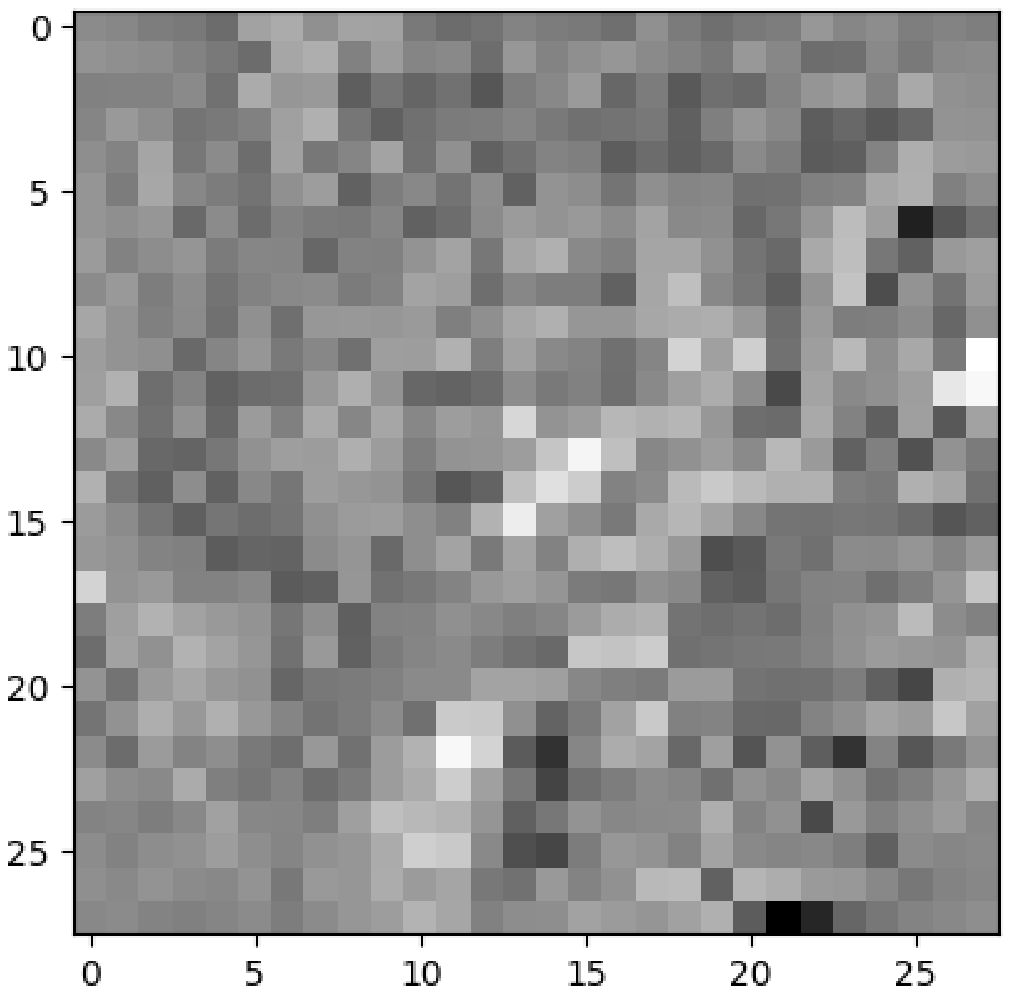}
	}
	\subfigure[\small Inverted ]
	{
          \includegraphics[width=.08\textwidth]{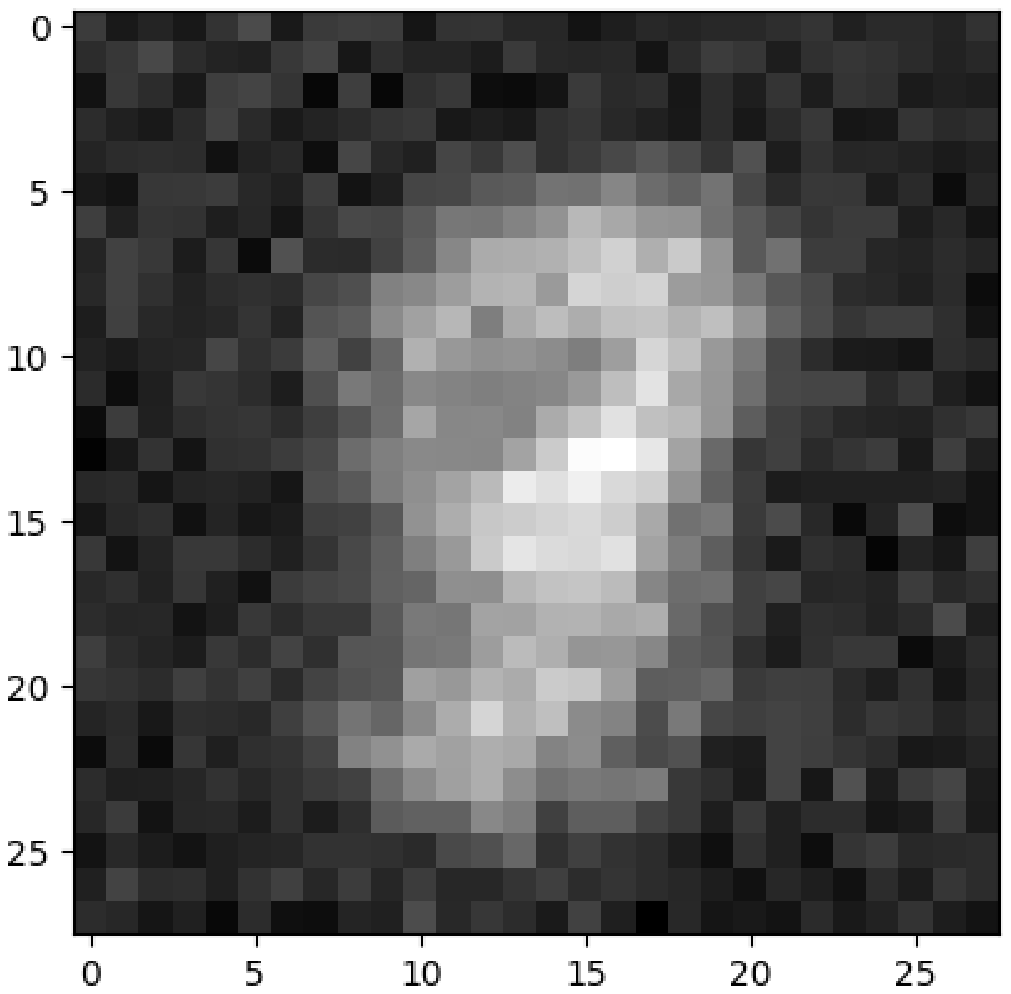}
	}
	\subfigure[\small Masked ]
	{
          \includegraphics[width=.08\textwidth]{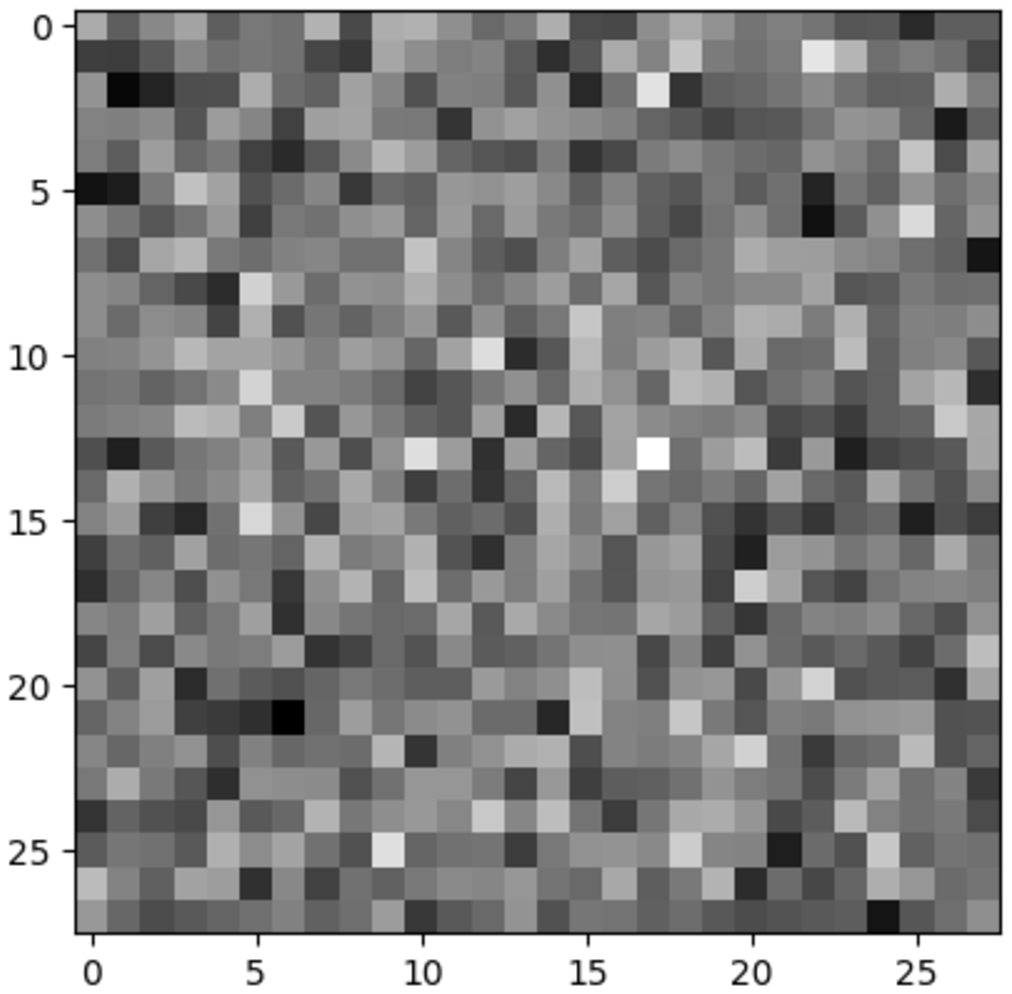}
	}
	\subfigure[\small Inverted ]
	{
          \includegraphics[width=.08\textwidth]{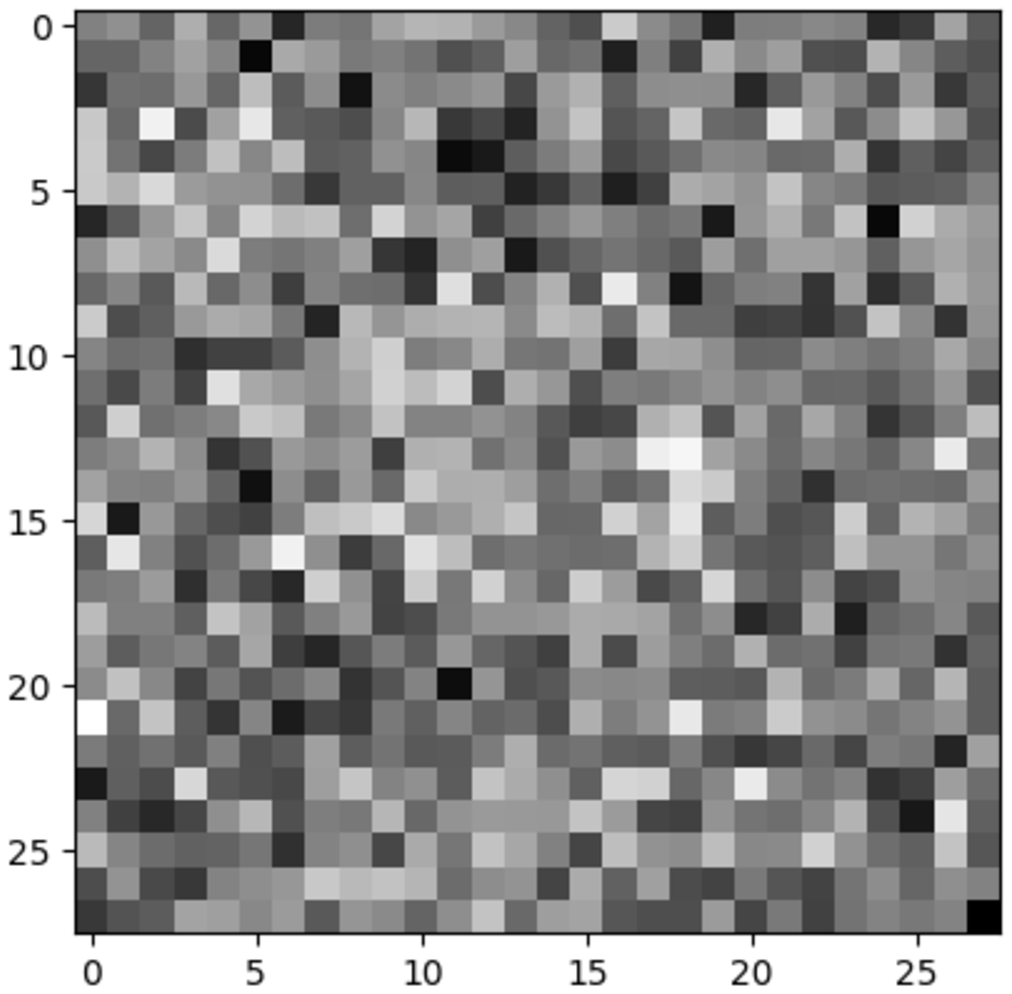}
	}

        \vspace{0em}
        \caption{Original, masked, reconstructed samples. (b) and (d) show the masked sample while (c) and (e) show ISP's reconstructions with the smallest MSEs when adversarial learning factor $\lambda$ is 0 and 0.6, respectively.}
	\label{fig:mnist-examples}
        \vspace{0em}
      \end{figure}
      
{\em Resilience against mobile-ISP collusion:} For this MNIST example, we only consider the privacy threat of inversion attack. Suppose one of the 100 mobiles colludes with ISP.
We measure the MSEs and SSIMs of the inversion attack on the non-colluding mobiles. MSE and SSIM are complementary in characterizing the privacy loss caused by the inversion attack. MSE measures the average pixel-wise difference between the original and reconstructed samples. However, MSE falls short of characterizing their correlation. SSIM, which is a perceptual metric quantifying quality degradation of reconstructed image, captures the correlation. Fig.~\ref{fig:reconstructioncdf_mnist} shows the CDFs of the MSEs when $\lambda=0$ and $\lambda=0.6$.
The vertical line in Fig.~\ref{fig:reconstructioncdf_mnist} shows the inversion MSE for the colluding mobile when $\lambda=0$, which is 0.71. The inversion MSEs for the non-colluding mobiles are distributed from 0.76 to 1.25, larger than that for the colluding mobile. When adversarial learning is not adopted (i.e., $\lambda=0$), the inversion MSEs for all the non-colluding mobiles are higher than that of the colluding mobile. When adversarial learning is adopted with $\lambda=0.6$, the inversion MSEs of the non-colluding mobiles are dispersed in a much wider range of 1.77 to 2,630, larger than the MSEs when $\lambda=0$.
Fig.~\ref{fig:reconstructioncdf_mnist_ssim} shows the CDFs of SSIMs when $\lambda=0$ and $\lambda=0.6$. Note that the maximum value of SSIM is 1, indicating the highest structural similarity.
From Fig.~\ref{fig:reconstructioncdf_mnist_ssim}, the SSIMs when $\lambda=0.6$ are smaller than those when $\lambda=0$, suggesting that adversarial learning reduces the structural similarity.
Fig.~\ref{fig:mnist-examples} shows the original and masked samples and the ISP's inversion results with the smallest MSE for a non-colluding mobile when $\lambda=0$ and $\lambda=0.6$. These samples show that adversarial learning is effective in counteracting the collusion-based inversion attack.


{\blue {\em Comparison with PAN \cite{liu2019privacy}:} As discussed in \sect\ref{sec:related}, the existing neural network masking approaches \cite{osia2017hybrid,wang2018not,liu2019privacy,li2021deepobfuscator,chi2018privacy,li2020tiprdc} have two basic differences from PriMask. First, they only address external eavesdroppers and do not consider curious ISP. Second, they are not cascadable since they apply custom designs for the InferNets. In this paper, we compare PriMask with Privacy Adversarial Network (PAN) \cite{liu2019privacy} when the ISP is curious and colludes with a mobile.  To enable the comparison, we set PAN's encoder's output dimension equal to the dimension of the MNIST samples. Table~\ref{tab:comparison1} shows the the statistics of test accuracy and median MSE achieved by the inversion attack under PAN and PriMask over 10 tests. The median MSEs achieved by the attack are 0.50 and 1.77 under PAN and PriMask, respectively. This suggests that, under PAN, the curious ISP can better reconstruct the inference samples. The test accuracy ranges under PAN and PriMask are similar. These results show that PAN provides weaker privacy protection. In \sect\ref{subsubsec:comparison}, we further compare PAN and PriMask under the private attribution extraction attack. }

\begin{table}
	\centering
	\caption{Comparison between PAN \cite{liu2019privacy} and PriMask.}
	\vspace{0em}
	\label{tab:comparison1}
	\small
	\begin{tabular}{ccccc}
		\toprule
		\multirow{2}{*}{}  & Test accuracy & Mean & Median MSE \\
		& percentiles (10\%,90\%) & test accuracy& of AttackNet\\
		\midrule
		PAN &  (0.92,0.99) & 0.96& 0.50\\
		PriMask &  (0.92,0.98) & 0.95 & 1.77 \\
		\bottomrule
	\end{tabular}
	\vspace{0em}
\end{table}

\begin{table}
	\centering
	\caption{Setting of $\lambda$.}
	\vspace{-1em}
	\label{tab:lambda}
	\normalsize
	\begin{tabular}{ccccc}
		\toprule
		$\lambda$ & Mean test accuracy& Median MSE of AttackNet\\
		\midrule
		0.2 & 0.96& 1.62\\
		0.4 & 0.95 & 1.67 \\
		0.6 & 0.95& 1.77\\
		0.8& 0.94 & 1.89 \\
		\bottomrule
	\end{tabular}
\end{table}

{\green {\em Setting of $\lambda$}: As discussed in \sect4.3, the PSP can perform SAL with different $\lambda$ settings and validate them using a small validation dataset. Table~\ref{tab:lambda} shows the $\lambda$ setting versus test accuracy and MSE of the inversion attack. We only show the results when $\lambda = 0.2,0.4,0.6,0.8$ since the training of adversarial learning is sensitive to the combination weights $\lambda$ of multiple losses \cite{groenendijk2021multi}. With the increase of $\lambda$, the mean test accuracy drops moderately while the mean MSE of AttackNet increases slightly. Note that with different $\lambda$ settings, the converge speeds are different. The results shows that the PSP can perform SAL for different $\lambda$ settings for heterogeneous users with different privacy budgets.}

{\blue  }

\section{Human Activity Recognition}
\label{sec:case1}

Human activity recognition (HAR) with the data from the inertial measurement units (IMUs) of a user's mobile is a basic building block of mobile sensing applications. However, IMU data may contain private information related to identity, gender, and age \cite{lu2013human,garofalo2019systematic}. In this section, we apply PriMask to an HAR system to counteract both the inversion attack and private attribute extraction.

\subsection{HAR Dataset, InferNet, and HyperNet}

We use a public dataset \cite{anguita2013public} collected from 30 human volunteers performing six types of daily activities (walking, walking upstairs, walking downstairs, sitting, standing, and laying).
Each volunteer carried a waist-mounted smartphone for recording the accelerometer and gyroscope data. The recorded data include 3-axial linear acceleration with/without gravity and 3-axial angular velocity sampled at $50\,\text{sps}$. Thus, each record has nine components. The record traces are pre-processed by noise filters and then arranged in sliding windows of 2.56 seconds with 50\% overlap. The trace within a window is referred to as a {\em data sample}. Thus, each data sample is a tensor sized $9 \times 1 \times 128$. Each data sample has an activity label. The dataset contains 10,299 data samples that are partitioned into the training and testing subsets by 7:3. Each data sample also has a volunteer identity label to indicate which volunteer that the sample was collected from. We regard the identity as the private attribute.



We design a CNN InferNet,
consisting of two convolutional layers with max pooling, three dense layers with ReLU activation, and softmax function. The first convolutional layer admits a 1,152-dimensional vector flattened from the data sample tensor and applies 32 $1 \times 9$ convolution filters. The second convolutional layer applies 64 $1 \times 9$ filters.
The three dense layers have 1,000, 500, and 6 neurons.
To avoid overfitting, we adopt dropout on dense layers. The test accuracy of the trained InferNet on raw data is 92.5\%.


The MaskNet adopts a two-layer MLP architecture. For both the input and output layers, the number of neurons is 1,152. The middle layer has 200 neurons. There are 230,600 trainable parameters between the input and middle layers, and 231,552 between the middle and output layers.
The HyperNet adopts a similar architecture as shown in Fig.~\ref{fig:mnisthyper}, with modifications on the number of neurons.

\subsection{Evaluation Results for HAR}

We train three HyperNets. For the first HyperNet, adversarial learning is disabled (i.e., $\lambda=0$). Therefore, this HyperNet is agnostic to the type of privacy attack. The other two HyperNets are adversarially trained to counteract the inversion attack (with $\lambda=0.3$) and private attribute extraction (with $\lambda=0.1$), respectively.
By default, we use each HyperNet to generate 100 MaskNets for evaluation. We also generate 100,000 MaskNets to evaluate scalability of PriMask. {\green Note that the scalability in this paper is in terms of the number of the generated MaskNets. The computational scalability of the ISP depends on the hardware specification of the ISP server and the concurrency of the service requests. The ISP can invest sufficient computational capacity to meet the demand.}


\subsubsection{Impact of PriMask on InferNet accuracy}
\label{subsubsec:har-primask-accuracy}

\begin{figure}
	\subfigure[\small CDF of accuracy]{	\includegraphics[width=0.146\textwidth]{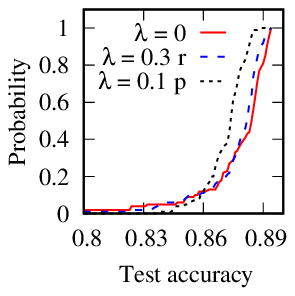}
		\label{fig:testacccdf_har}
	}
	\subfigure[\small CDF of MSE]{
		\includegraphics[width=0.146\textwidth]{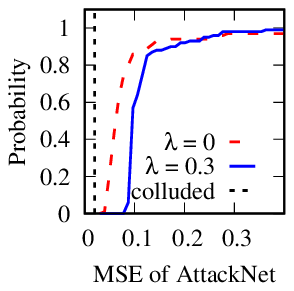}
		\label{fig:reconstructioncdf_har}
	}
	\subfigure[\small CDF of ASR]{
		\includegraphics[width=0.146\textwidth]{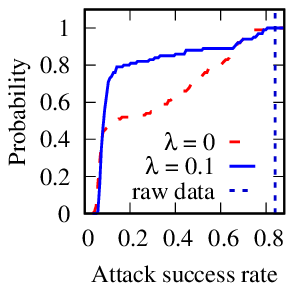}
		\label{fig:privatecdf_har}
              }
              \vspace{0em}
              \caption{Impact of PriMask on HAR test accuracy and privacy protection. Legends denoted by `r' and `p' are for HyperNets adversarially trained with inversion attack and private attribute extraction, respectively.}
              \vspace{-1em}
	\label{fig:cdfs_har}	
\end{figure}



Fig.~\ref{fig:testacccdf_har} shows the CDF of the InferNet's test accuracy corresponding to the three HyperNets. The test accuracies are distributed within $(80\%, 90\%)$. The average test accuracies for the three CDFs are 87.6\%, 87.0\%, 87.5\%. Thus, on average, there are accuracy drops of 5\% to 5.5\%.

\begin{table}
	\centering
	\caption{Statistics of InferNet's test accuracies across 100,000 MaskNets ($\lambda=0.1$; AttackNet = ExtNet).}
	\vspace{-1em}
	\label{tab:har-massive-testacc}
	\begin{tabular}{cccc}
		\toprule
		Post-generation validation & test accuracy range & mean \\
		\midrule
		Not applied  &  (0.52, 0.90) & 0.87 \\
		Applied &  (0.80, 0.90) & 0.88 \\
		\bottomrule
	\end{tabular}
	\vspace{0em}
\end{table}

We further evaluate PriMask's scalability in terms of InferNet accuracy. We generate 100,000 MaskNets using the HyperNet adversarially trained against private attribute extraction with $\lambda=0.1$.
The first row of Table~\ref{tab:har-massive-testacc} summarizes InferNet's test accuracies across all MaskNets. The mean value of the test accuracies (i.e., 87\%) remains at the same level as in Fig.~\ref{fig:testacccdf_har}. This result is supportive of PriMask's scalability in terms of InferNet accuracy. However, a few MaskNets lead to low InferNet accuracy of down to 52\%. Less than 2\% of all the MaskNets lead to InferNet accuracy lower than 80\%. This suggests a long-tail distribution of InferNet's accuracies across the MaskNets. To avoid outlier MaskNets, we apply a post-generation validation process. Specifically, for each generated MaskNet, the PSP uses a small validation dataset and works with the ISP to measure the InferNet's test accuracy. The PSP regenerates the MaskNet until the InferNet's test accuracy exceeds a {\em passing threshold}. Only the validated MaskNets are released. The second row of Table~\ref{tab:har-massive-testacc} summarizes the results if the validation is applied, where PSP's validation dataset includes 100 samples and passing threshold is 80\%. The average test accuracy increases to 88\%.

\begin{figure}
	\subfigure[\small HyperNet trained without adversarial learning, i.e., $\lambda = 0$]
	{
          \includegraphics{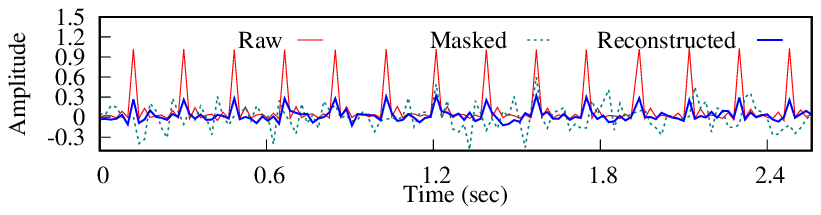}
          \label{fig:example_har_ori}
	}
	\subfigure[\small HyperNet trained against inversion attack ($\lambda=0.3$)]
	{
          \includegraphics{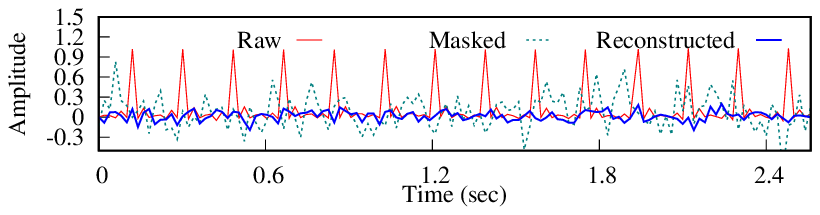}
          \label{fig:example_har_adv_rec}
	}
        \subfigure[\small HyperNet trained against private attribute extraction ($\lambda = 0.1$)]
	{
          \includegraphics{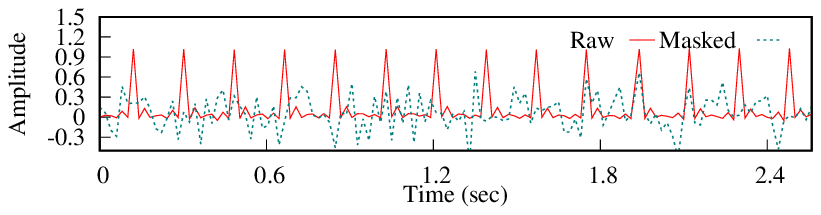}
          \label{fig:example_har_adv_pri}
	}
        \vspace{0em}
	\caption{Traces of raw and ISP's reconstructed data for the first axis of linear acceleration, as well as the first dimension of masked data for a certain non-colluding mobile. The ground truth human activity is standing.}
	\label{fig:har-examples}
        \vspace{-1em}
      \end{figure}

\subsubsection{Resilience against a single colluding mobile}
\label{sec:har-single}

We consider a system of 100 mobiles and one of them colludes with ISP.
Fig.~\ref{fig:reconstructioncdf_har} shows the CDFs of MSEs achieved by InvNet for non-colluding mobiles when the HyperNet is trained without or with adversarial learning. The vertical line represents the inversion MSE (i.e., 0.02) for the colluding mobile when $\lambda=0$. The MSEs for non-colluding mobiles are higher than that for the colluding mobile. In addition, when adversarial learning is applied, MSEs are larger.

Fig.~\ref{fig:privatecdf_har} shows the CDF of the attack success rate (ASR), i.e., the accuracy of the extracted private attribute, achieved by ExtNet on raw IMU data and masked data from non-colluding mobiles. The two CDFs are results for the HyperNets with and without adversarial learning against ExtNet. The vertical line represents ASR on raw IMU data (i.e., 84\%). Note that since the data samples are collected from 30 volunteers \cite{anguita2013public}, the random guessing strategy yields an ASR of $1/30=3.3\%$. The 84\% ASR suggests that the IMU data contains abundant information regarding user identity. When adversarial learning is applied, the CDF is higher than that without adversarial learning. This shows that adversarial learning is effective in reducing ASR. The average ASRs with and without adversarial learning are $28.3\%$ and $17\%$, respectively.
Although MaskNets under the setting of $\lambda=0.1$ cannot reduce ASR to the random guessing level, they have already achieved significant reductions in ASR compared with the case without private attribute protection. By setting larger $\lambda$, the ASRs for non-colluding mobiles will further decrease. But InferNet's accuracy will decrease too.

The three subfigures of Fig.~\ref{fig:har-examples} show the traces of the raw and ISP's reconstructed data for the first axis of linear acceleration, and the first dimension of the masked data, for a certain non-colluding mobile adopting the three HyperNets, respectively. For linear acceleration, the peaks are salient features. From Fig.~\ref{fig:har-examples}, the masked traces do not have salient peaks. In Fig.~\ref{fig:example_har_ori}, without adversarial learning, the reconstructed trace still pronounces peaks at the same times of the original peaks. In Fig.~\ref{fig:example_har_adv_rec}, with adversarial learning, the reconstructed trace no longer pronounces peaks at the same times of the original peaks, suggesting better protection against inversion attack.

\begin{figure}
  \subfigure[\small ASR vs. ExtNet]{	\includegraphics[width=0.145\textwidth]{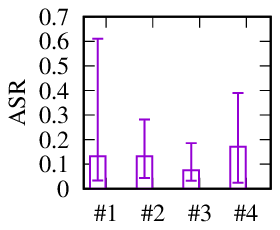}
    \label{fig:example_har_different_extnet}
  }
  \subfigure[\small ASR vs. \# of colluders]{
    \includegraphics[width=0.305\textwidth]{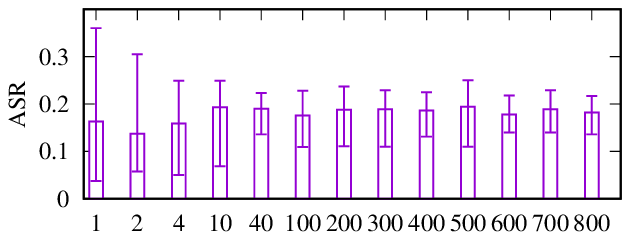}
    \label{fig:example_har_errorbar_attack}
  }
  \vspace{0em}
  \caption{Impact of ISP's ExtNet architecture and number of colluding mobiles on ASR. In (a), ExtNet\#1 is also the architecture used by PSP in SAL. Whiskers of an error bar represent maximum and minimum.}
  \label{fig:example_har_additional_asrs}
  \vspace{0em}
\end{figure}

Next, we investigate the impact of the ExtNet architecture used by ISP on ASR of private attribute extraction. We consider a system of 1,000 mobiles and one of them colludes with ISP. The HyperNet adversarially trained against ExtNet with $\lambda=0.1$ is used to generate MaskNets. We design four ExtNet architectures used by ISP, which are illustrated as:

$\bullet$ {\em ExtNet\#1:} $\text{C}_{32}$-$\text{C}_{64}$-$\text{D}_{1664}$-$\text{D}_{1000}$-$\text{D}_{500}$-$\text{D}_{30}$-softmax

$\bullet$ {\em ExtNet\#2:} $\text{C}_{32}$-$\text{C}_{64}$-$\text{C}_{128}$-$\text{D}_{1152}$-$\text{D}_{500}$-$\text{D}_{30}$-softmax

$\bullet$ {\em ExtNet\#3:} $\text{D}_{1152}$-$\text{D}_{100}$-$\text{D}_{50}$-$\text{D}_{30}$-softmax

$\bullet$ {\em ExtNet\#4:} $\text{D}_{1152}$-$\text{D}_{500}$-$\text{D}_{100}$-$\text{D}_{30}$-softmax

\noindent where $\text{C}_n$ represents a convolutional layer with $n$ filters followed by max pooling, $\text{D}_{n}$ represents a dense layer of $n$ neurons with ReLU activation. The ExtNet\#1 architecture is identical to that used by the PSP's adversarial learning.
In Fig.~\ref{fig:example_har_different_extnet}, each error bar shows the average, maximum, minimum of the ASRs across the 999 non-colluding mobiles when ISP adopts a certain ExtNet architecture. No ExtNet architecture shows clear advantage for ISP in terms of average ASR.
This suggests that the resilience against collusion is insensitive to the ExtNet architecture used by the ISP.

\subsubsection{Resilience against multiple colluding mobiles}

We consider a system of 1,000 mobiles and vary the number of mobiles colluding with ISP.
For a certain set of colluding mobiles, to build the training dataset for constructing ExtNet, ISP feeds each original training sample to all colluding mobiles' MaskNets to obtain multiple samples for training ExtNet. Thus, the ISP's trained ExtNet is expected to address all colluding mobiles' MaskNets. Fig.~\ref{fig:example_har_errorbar_attack} shows the error bars of non-colluding mobiles' ASRs versus the number of colluding mobiles. While the average ASR exhibits an increasing trend when the number of colluding mobiles is lower than 10, it becomes flat at around 20\% when the number of colluding mobiles is up to 800. Recall that, without PriMask's protection, ASR is up to 84\% (cf.~\sect\ref{sec:har-single}). The above results suggest that PriMask is resilient to increase of colluding mobiles. Note that the range of ASR shows a decreasing trend with the number of colluding mobiles. This is partly due to decreasing number of non-colluding mobiles for generating the ASR statistics (i.e., minimum/maximum).

\subsubsection{Comparison with PAN}
\label{subsubsec:comparison}

\begin{figure}
	\subfigure[\small Error bar of accuracy]{	
		\includegraphics[width=0.144\textwidth]{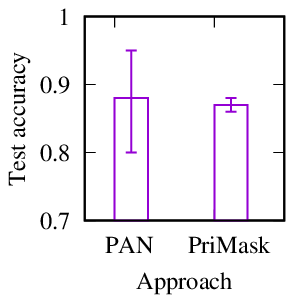}
		\label{fig:comparison1}
	}
	\subfigure[\small Error bar of ASR]{
		\includegraphics[width=0.144\textwidth]{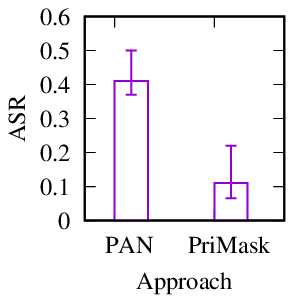}
		\label{fig:comparison2}
	}	
	\subfigure[\green \small ASR vs accuracy]{
	\includegraphics[width=0.144\textwidth]{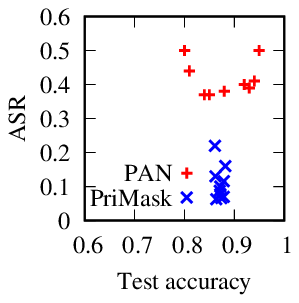}
	\label{fig:comparison3}
	}
	\vspace{0em}
	\caption{Comparison between PAN \cite{liu2019privacy} and PriMask.}
	\label{fig:comparison_all}
	\vspace{0em}
\end{figure}


{\blue In this experiment, we compare the resilience of PAN and PriMask against the private attribute extraction attack. Similar to \sect\ref{subsec:mnist}, to enable the comparison, we adjust the encoder structure of PAN \cite{liu2019privacy} such that the size of the encoded features equals the dimension of the raw data sample. Specifically, we design an encoder with two convolutional layers and the output size of 1152. We adopt a two-layer MLP for inference. ExtNet in PAN also adopts a similar two-layer MLP architecture.  Fig.~\ref{fig:comparison_all} shows the error bars of InferNet accuracy and ASR, as well as  InferNet accuracy versus ASR under PAN and PriMask {\green (privacy-utility trade-off points)}. For PAN, each data point corresponds to a setting of the Lagrangian multiplier $\lambda_p$, which affects the privacy protection level of PAN \cite{liu2019privacy}. For PriMask, the trade-off data points correspond to the HyperNet-generated MaskNets. {\green The trade-off points show that with the enhancement of privacy protection, the test accuracy under both PAN and PriMask drops gradually. However, }from the results, PriMask achieves better privacy protection than PAN, subject to similar InferNet test accuracy.

 }

%
%
%

\section{Urban Environment Crowdsensing}
\label{sec:case2}


We study the data \cite{cao2018walkway} collected in a city-wide experiment that involves over 10,000 school students in our city to understand urban environment in a crowdsensing manner. 
 In the experiment, each participant carries a wearable device on a neck lanyard which integrates several sensors to record the surrounding environment conditions of the participant. The sensor measurements are transmitted opportunistically to a central cloud portal through over 18,000 Wi-Fi hotspots deployed across our city. On this crowdsensing platform, we develop an urban environment crowdsensing (UEC) application that classifies the ambient conditions of participants and identifies those which may result in discomfort for human beings. We investigate the potential of location leakage and apply PriMask to protect participants’ location privacy in such a circumstance.

\subsection{Dataset, InferNet, ExtNet, and PriMask}

The dataset consists of 640,000 samples. Each sample includes five sensor readings of light intensity, noise level, atmospheric pressure, ambient temperature, and body-reflected temperature. Each sample is geotagged using location information inferred from the wearable device's nearby Wi-Fi hotspots. To build the UEC application, we only use the five sensor readings. To investigate location privacy leakage, we study the correlation between the sensor readings and the geotags. The city-wide experiment collected geotags for ground truth only. After the privacy-preserving UEC application is developed and deployed, geotags are no longer needed.



The wearable device used in the experiment does not support the participants to key in their real-time comfort feelings. In fact, requesting the participants to continuously or frequently provide feedback is impractical. Thus, we follow several first principles to generate the ground truth information regarding the discomfort level of the environment condition. The procedure is as follows. First, we normalize each sensor reading to $[0, 1]$.
Then, we apply the following three scoring functions on the normalized sensor readings to generate discomfort scores: $d_1(x) = x$, $d_2(x) = 1-x $, and $d_3(x) = 4 \cdot (x - 0.5)^2$.
As human discomfort in general increases with noise level, we apply $d_1(x)$ to score noise level.
In the climate zone of our city, as low atmospheric pressure is in general positively correlated with human discomfort, we apply $d_2(x)$ to score atmospheric pressure. As the light intensity, ambient temperature, and body-reflected temperature should be in their respective proper ranges, we apply the quadratic function $d_3(x)$ with the minimum (i.e., the least discomfort) at $x=0.5$ on them.
Lastly, we sum up the five scores to obtain a final score to characterize the overall human discomfort. The users with high discomfort scores may need attention and preventative actions.

We generate the private attribute labels as follows.
First, we apply the $k$-means algorithm with $k=10$ to cluster the data's geotags into ten zones.
We view the zone ID as the private attribute. However, the numbers of samples in the clusters are imbalanced. To simplify the presentation of the evaluation results, we re-sample the data to ensure class balance. The re-sampling generates a training dataset of 60,000 samples (i.e., 6,000 samples in each zone) and a testing dataset of 4,481 samples (i.e., about 448 samples in each zone).



Transmitting the participants' data to the cloud portal is preferred, because it supports various posterior data analytics including UEC. If raw data is transmitted, the scoring functions can be applied directly. However, to admit masked data, a regression InferNet is needed to approximate the sum of the scoring functions. We design an MLP with three hidden layers as the InferNet, which have 10, 20, and 10 neurons with ReLU activation.
The output layer is a single neuron giving the predicted discomfort score. We still use the test accuracy to characterize the performance of the InferNet. Specifically, if the difference between the predicted score and the ground truth is less than 0.5 (i.e., 10\% of the maximum discomfort score), we regard the prediction correct. The trained InferNet achieves a test accuracy of 99.2\% on the raw test data.

Then, we train an ExtNet using the training data and the associated zone IDs. The ExtNet has two hidden layers, with 200 and 50 neurons using ReLU activation, respectively. It has an output layer with 10 neurons corresponding to the 10 zones. On the raw testing data, the ExtNet's ASR is 42\%. As the strategy of randomly guessing the zone has 10\% ASR only, the ExtNet's 42\% ASR suggests that the sensor readings leak information regarding the participants' locations.

The MaskNet is an MLP with a single hidden layer. The input, hidden, and output layers have five neurons each. The MaskNet has 60 trainable parameters.
The HyperNet adopts a similar architecture as shown in Fig.~\ref{fig:mnisthyper}, with minor modifications on the number of neurons for MaskNet compatibility.

\subsection{Evaluation Results for UEC}
\begin{figure}
	\subfigure[\small CDF of accuracy]{	\includegraphics[width=0.146\textwidth]{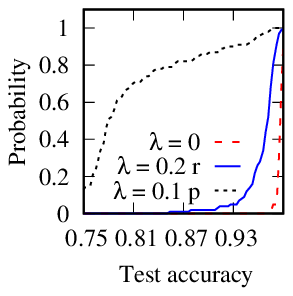}
		\label{fig:testacccdf_nse}
	}
	\subfigure[\small CDF of MSE]{
		\includegraphics[width=0.146\textwidth]{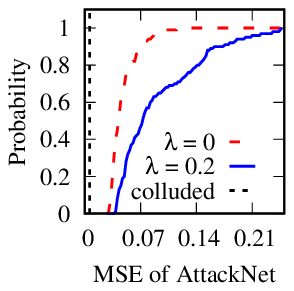}
		\label{fig:reconstructioncdf_nse}
	}
	\subfigure[\small CDF of ASR]{
		\includegraphics[width=0.146\textwidth]{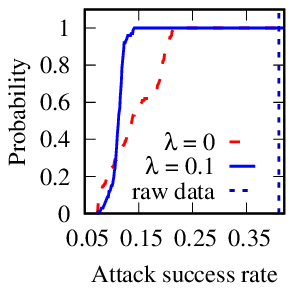}
		\label{fig:privatecdf_nse}
	}
	\vspace{-1em}
	\caption{Impact of PriMask on UEC test accuracy and privacy protection. Legends denoted by `r' and `p' are for HyperNets adversarially trained with inversion attack and privacy extraction.}
	\label{fig:cdfs}
	\vspace{-1em}
\end{figure}
We train three HyperNets with the following settings: 1) $\lambda=0$; 2) $\lambda=0.2$ against inversion attack; 3) $\lambda=0.1$ against private attribute extraction.
Fig.~\ref{fig:testacccdf_nse} shows the CDFs of the test accuracies when the three HyperNets are used. The average test accuracy for the first two HyperNets are 98.3\% and 96.2\%, respectively. Thus, compared with the test accuracy obtained on raw test data (i.e., 99.2\%), there are 0.9\% and 3\% accuracy drops. When the third HyperNet is used, the average test accuracy drops to 80.1\%. This is because $\lambda=0.1$ is an aggressive setting against private attribute extraction. We will explain this issue along with the achieved privacy protection strength shortly.


Suppose a mobile colludes with ISP.
Fig.~\ref{fig:reconstructioncdf_nse} shows the CDFs of inversion MSEs when the first and the second HyperNets are used.
The adversarial learning improves resilience against inversion attack in the presence of collusion. Fig.~\ref{fig:privatecdf_nse} shows the CDFs of ASRs when the first and the third HyperNets are used.
When no adversarial learning (i.e., $\lambda=0$) is applied, PriMask reduces ASR from the original 42\% to 14.2\% on average. As the 14.2\% ASR is close to its lower bound of 10\%, intuitively, more data utility will be sacrificed to further reduce ASR. Thus, when $\lambda=0.1$ that produces an average ASR of 11\%, we observe significant test accuracy drops in Fig.~\ref{fig:testacccdf_nse}. Smaller $\lambda$ settings can restore the test accuracy, while the resulting average ASRs will be within $(11\%, 14.2\%)$, which are satisfactory.

\section{Driver Behavior Recognition}
\label{sec:case3}

In U.S., one in five car accidents is caused by a distracted driver \cite{distracted-driving}. Thus, using smartphone to detect driver's engagement in distracted behaviors is useful. To incentivize drivers' participation, the car insurance companies may provide premium discounts according to the monitoring results.
To facilitate the design of driver behavior recognition (DBR), an insurance company initiated a competition \cite{driver2021} by providing a dataset of images captured in cars regarding drivers' behaviors. In this section, we train a CNN based on the dataset. From our implementation, the CNN is heavy ($226\,\text{MB}$) and inappropriate for local execution on phones. If we run it in the cloud, transmitting the raw images to the cloud inevitably incurs privacy concerns. Thus, in this section, we apply PriMask to design privacy-preserving cloud-based DBR.

\subsection{DBR Dataset and System Design}

\begin{figure}
	\subfigure[\small CDF of accuracy]{	\includegraphics[width=0.146\textwidth]{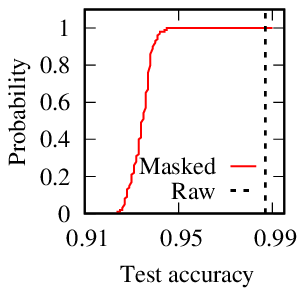}
		\label{fig:testacccdf_driver}
	}
	\subfigure[\small CDF of MSE]{
		\includegraphics[width=0.146\textwidth]{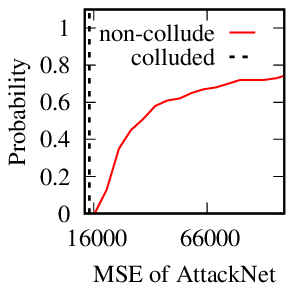}
		\label{fig:reconstructioncdf_driver}
	}
	\subfigure[\small CDF of ASR]{
		\includegraphics[width=0.146\textwidth]{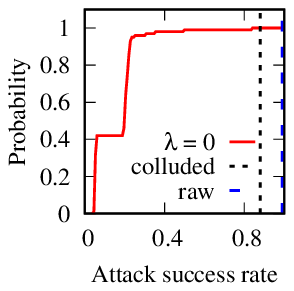}
		\label{fig:privatecdf_driver}
	}
	\vspace{0em}
	\caption{Impact of PriMask on DBR accuracy \& privacy.}
	\label{fig:cdfs_driver}
	\vspace{0em}
\end{figure}

\begin{figure}
	\subfigure[\small Original]
	{
		\includegraphics[width=0.22\columnwidth]{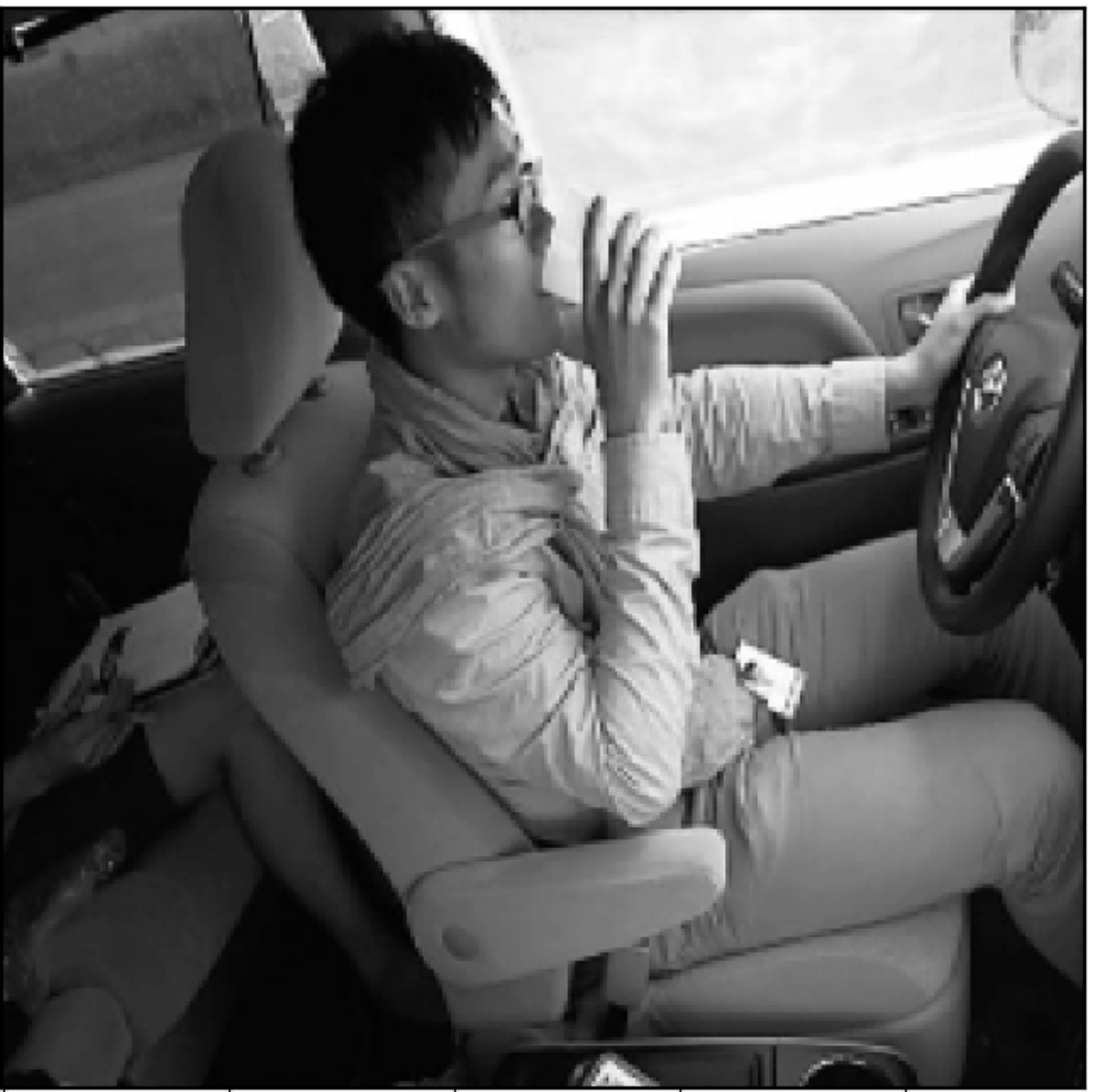}
	}
	\subfigure[\small Masked]
	{
		\includegraphics[width=0.22\columnwidth]{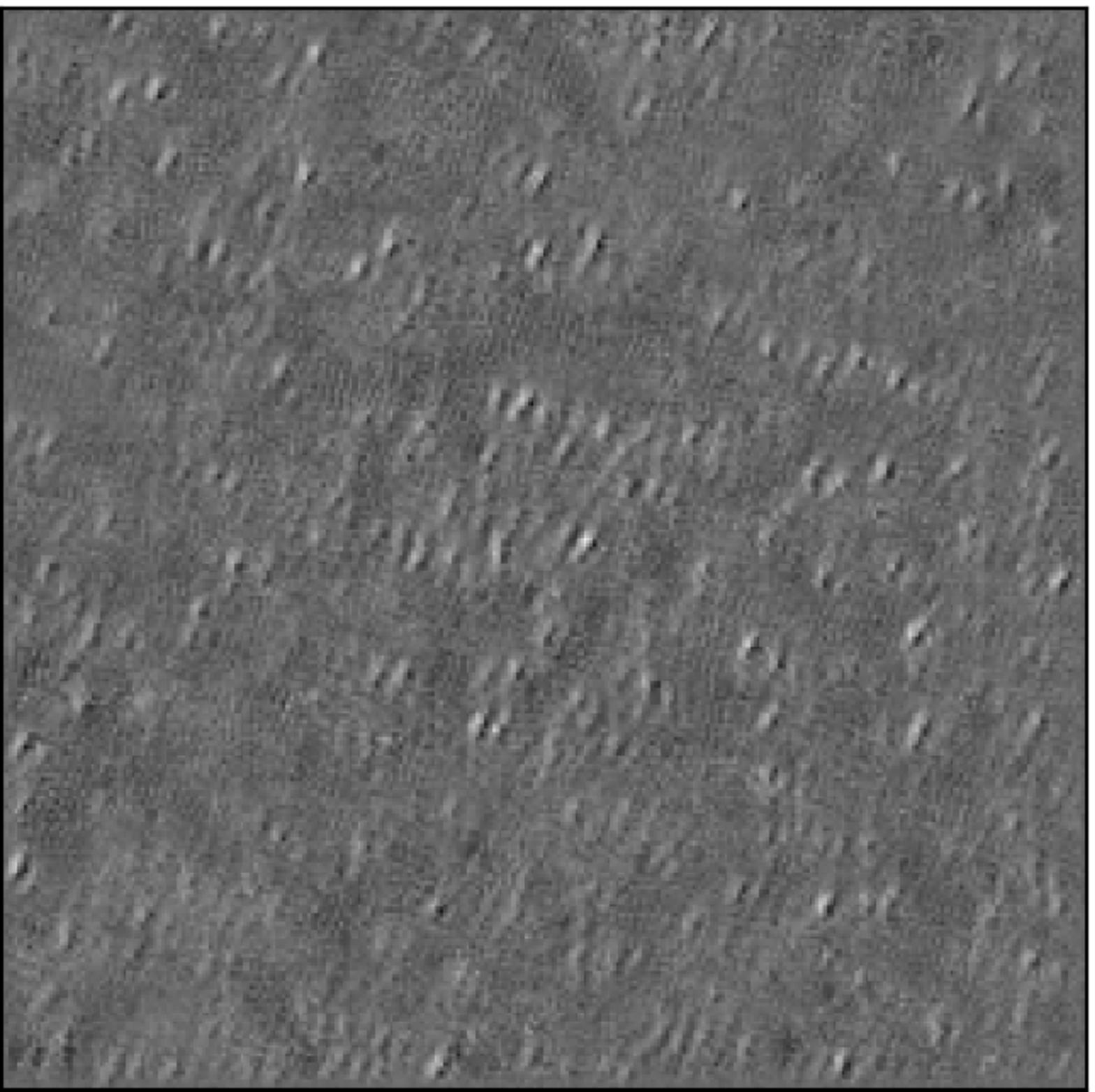}
	}
	\subfigure[\small Inversion]
	{
		\includegraphics[width=0.22\columnwidth]{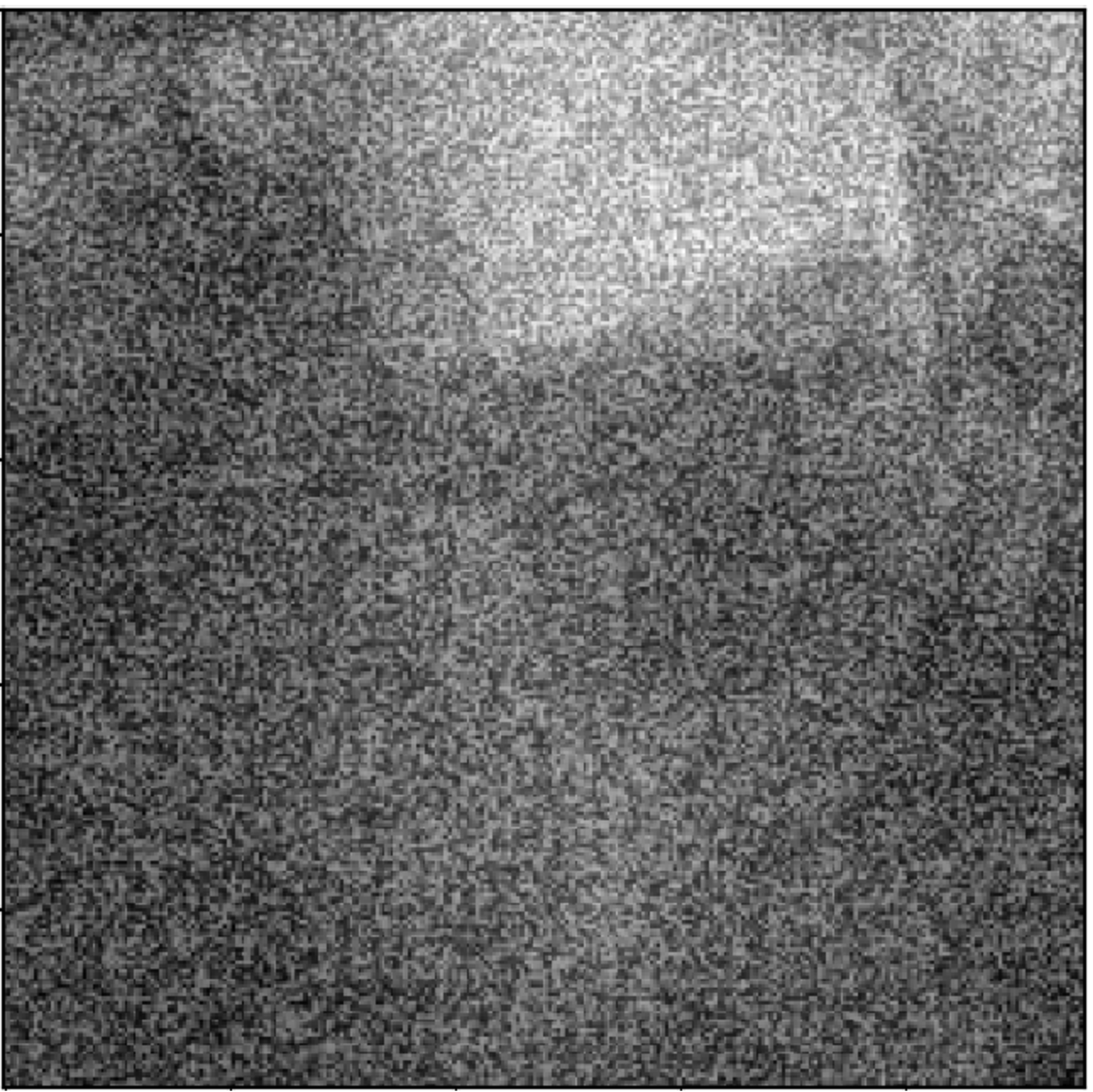}
	}
	\subfigure[\small Inversion]
	{
		\includegraphics[width=0.22\columnwidth]{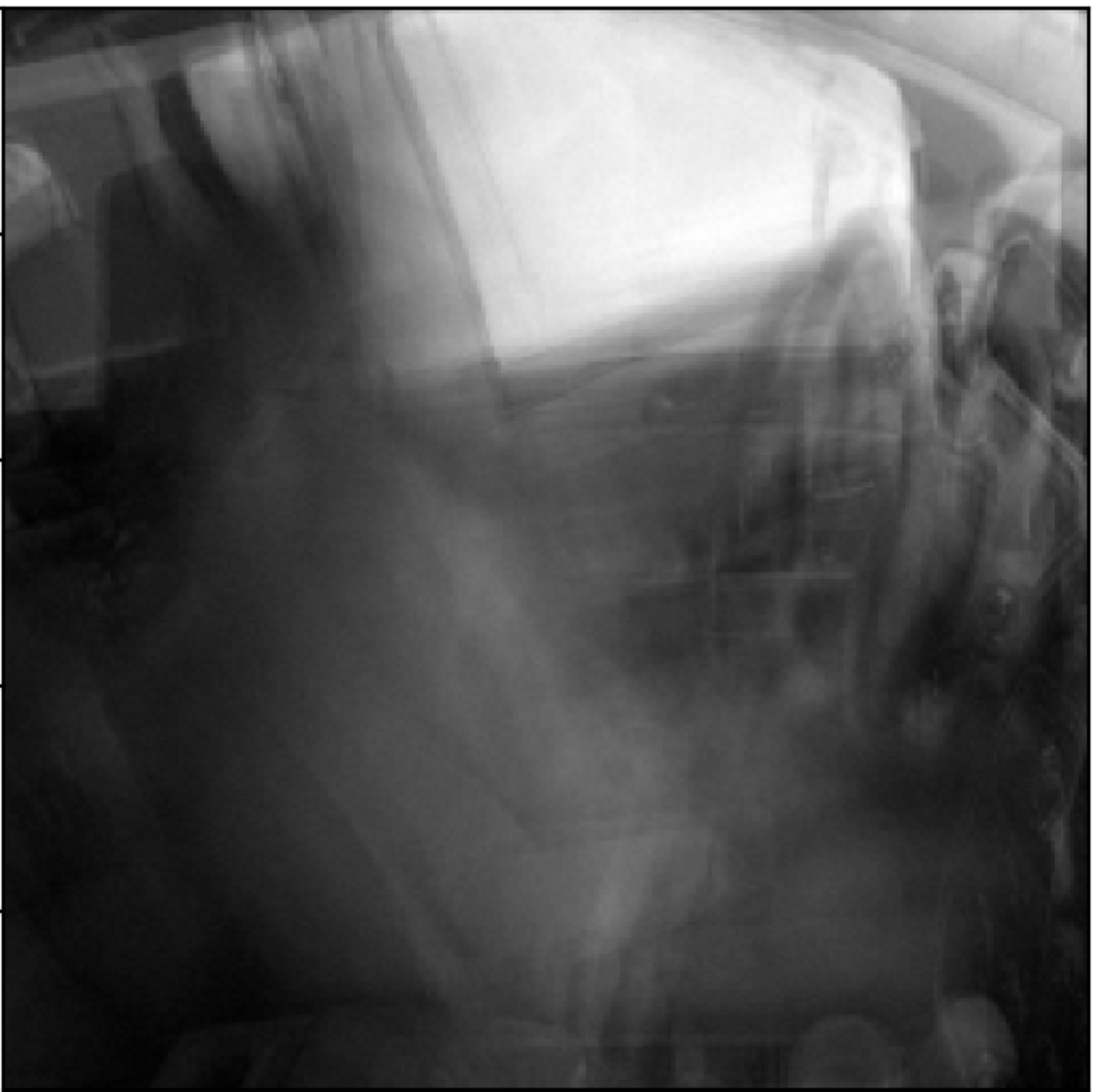}
	}
	\vspace{0em}
	\caption{Original, masked, reconstructed samples. The inversion MSE for (d) is the smallest among all non-colluding mobiles, i.e., this example is the worst case for non-colluding mobiles. Only {\em a priori} global information of the dataset (e.g., rough contours of car window and driver) can be seen from reconstructed samples, which are not specific to a certain driver and thus not private. In fact, ISP can generate an image similar to (d) by averaging all training samples.}
	\label{fig:driver-examples2}
	\vspace{0em}
\end{figure}

The dataset consists of 22,424 grayscale images, each sized $240 \times 240$. Each sample has a driver behavior label (in 10 classes) and driver identity as a private attribute label (in 26 classes). Examples of the driver behavior include safe driving, drinking, etc.
We partition the dataset into training samples and testing samples by 8:2.

DBR is a complex task. We implement a 32-layer CNN architecture described in \cite{distracted-driving-cnn}. Specifically, the CNN consists of 6 groups of convolutional, ReLU, and batch normalization, max pooling, and dropout layers, followed by 3 dense layers with ReLU, batch normalization, and dropout.
The CNN's test accuracy on raw testing samples is 98.46\%. It consists of more than 59 million parameters and requires $226\,\text{MB}$ memory space. Thus, this CNN is heavy for smartphones. Continuously running it on smartphone drains battery quickly. Thus, running it in the cloud is preferred. The overhead for the phone to transmit the images to the cloud is low. Assuming the phone records an image every five seconds and no image compression is applied, the phone only needs a bandwidth of $92\,\text{kbps}$ to sustain the transmission, which is little for today's broadband cellular connectivity.

The MaskNet uses a two-layer MLP architecture. For both the input and output layers, the number of neurons is 57,600. The middle layer has 120 neurons. The MaskNet has 13.9 million parameters and requires $53.61\,\text{MB}$ memory space. Thus, it is 7x smaller than InferNet in terms of memory usage. Moreover, MaskNet's dense layers require much less compute time than InferNet's convolutional layers. HyperNet architecture is similar to Fig.~\ref{fig:mnisthyper}, with minor changes on neuron numbers for MaskNet compatibility.

\subsection{Evaluation Results for DBR}

We train a HyperNet with $\lambda=0$.
As this HyperNet achieves good privacy preservation as shown shortly and adversarial learning often requires more training epochs, we omit adversarial learning.
Fig.~\ref{fig:testacccdf_driver} shows CDF of the test accuracies corresponding to all MaskNets. The average accuracy is 93.4\%. Compared with that on the raw data (i.e., 98.46\%), there is a drop of 5.1\% on average.

Suppose a mobile colludes with ISP. As shown in Fig.~\ref{fig:reconstructioncdf_driver}, the ISP achieves an inversion MSE of 14,069. This MSE is much larger than those seen for the MNIST example in \sect\ref{subsec:mnist}, because the MNIST and DBR samples have different pixel value ranges (i.e., $[0,1]$ vs. $[0,255]$). Fig.~\ref{fig:reconstructioncdf_driver} also shows the CDF of the inversion MSEs for the non-colluding mobiles. Such MSEs are distributed in a wide range from 16,429 to 556,000, with mean and median of 75,484 and 31,304, respectively.
Fig.~\ref{fig:driver-examples2} shows an original sample in subfigure (a), a non-colluding mobile's masked data in (b), and ISP's inversion results for two non-colluding mobiles in (c) and (d). The inversion MSEs of Fig.~\ref{fig:driver-examples2}(c) and (d) are 28,886 and 16,429, which are smaller than the average MSE and the smallest MSE among the non-colluding mobiles, respectively.
Thus, Fig.~\ref{fig:driver-examples2}(d) is the worst case for the non-colluding mobiles. However, we cannot observe useful information specific to the original sample in Fig.~\ref{fig:driver-examples2}(a). We can only observe {\em a priori} information that is applicable to the whole dataset, e.g., rough contours of a car window and a driver. Such {\em a priori} global information of the DBR application should not be viewed as a particular driver's privacy. In fact, the ISP can generate an image similar to Fig.~\ref{fig:driver-examples2}(d) by averaging all training samples.
The example shown in Fig.~\ref{fig:driver-examples2} suggests that the HyperNet achieves good privacy preservation against the inversion attack.

We also evaluate PriMask's resilience against private attribute extraction in the presence of collusion. ExtNet adopts the same architecture as InferNet, except that the last layer has 26 neurons corresponding to the volunteers. On raw data, ExtNet achieves 99.2\% ASR. Fig.~\ref{fig:privatecdf_driver} shows the ASRs based on raw data and the colluding mobile's masked data (i.e., 84.4\%), as well as CDF of ASRs for non-colluding mobiles. The mean ASR for non-colluding mobiles is 14.8\%. Thus, PriMask significantly reduces the attack effectiveness.









\section{Discussions}
\label{sec:discuss}

{\em Model and privacy leakage during SAL:} {\blue Although this paper assumes that the PSP is trustworthy, we discuss whether model leakage can occur between PSP and ISP. First, we consider whether the {\em static model extraction attacks} \cite{gonginversenet,kariyappa2021maze,miura2021megex} can be used by the PSP to exfoliate the ISP's InferNet. These attacks share a method of training a similar model without the private training data samples. However, the method requires that the attacker (i.e., the PSP in the current context) can query the static victim model  (i.e., InferNet in the current context) and obtain the predicted labels. As the ISP in SAL does not release the predicted labels and only renders the gradients of masked training samples to the PSP, the attack method is not directly applicable. Second, we consider whether the {\em inverted model extraction attack} against split learning studied in \cite{erdogan2021unsplit} can be used by ISP to extract the PSP's HyperNet. The study \cite{erdogan2021unsplit} shows the possibility for the server to extract the client's model part using a coordinate gradient descent approach under the split learning scheme. For PriMask, since the temporary MaskNets as the clients' model parts are heterogeneous and dynamically generated by the PSP, whether the attack described in \cite{erdogan2021unsplit} can exfoliate the HyperNet remains an open problem. From the above discussions, the issue of model leakage between the ISP and PSP requires further research for better understanding. There is also a recent study \cite{pasquini2021unleashing} showing the training data leakage vulnerability during split learning, as we have discussed in \sect\ref{sec:related}. The threat model in \cite{pasquini2021unleashing} in the context of this paper is as follows: the ISP refrains from transmitting training data to the PSP and the PSP is curious about the privacy contained in ISP's training data. Differently, as highlighted in \sect\ref{subsec:system model}, this paper considers a PSP trusted by the ISP and thus does not consider protecting the privacy contained in ISP's training data against a curious PSP. Advancing PriMask to address a PSP that is not trusted by ISP is an interesting topic for future work.}

{\em Composite privacy threats:} In this paper, we separately handle the inversion attack and private attribute extraction. The SAL method can be extended to jointly address multiple privacy attacks. Specifically, the composite loss function in Eq.~(\ref{eq:mixedlossrec}) can incorporate multiple attack losses (i.e., inversion MSE and ASRs regarding multiple private attributes). During the adversarial learning phase, an InvNet and multiple ExtNets can be trained against a temporary MaskNet. 


{\em Validated privacy protection:} In \sect\ref{subsubsec:har-primask-accuracy}, we presented a post-generation validation process to check the quality of a generated MaskNet in terms of InferNet accuracy. Similarly, we can also check in terms of privacy protection against potential mobile-ISP collusion. Now, we use the private attribution extraction as an example to discuss this. We say two MaskNets are {\em conflicting} if the collusion between any of them and the ISP leads to ASR against the other MaskNet higher than a passing threshold. The ASR can be measured using a validation dataset and the two MaskNets' respective ExtNets trained by the PSP. When generating the $(n+1)^\text{th}$ MaskNet, the PSP regenerates the candidate until it does not conflict with any of the previously released $n$ MaskNets. Now, we analyze the computation complexity of the above validation process. Assume that the probability that any two freshly generated MaskNets are conflicting is $p$ and the conflict statuses of any two MaskNet pairs are independent. Then, the expected number of generation processes needed for the $(n+1)^\text{th}$ MaskNet is $\frac{1}{(1-p)^n}$. Although the validation process is not scalable in general due to the exponential complexity, it can support a large enough system depending on the needed level of privacy protection. For instance, for the HAR application, the ASR on the raw IMU data is 84\%. By setting the ASR passing threshold to 58\%, the validation process enables the system with PriMask to provide privacy protection that is validated and better than the system without PriMask. From our measurements, the corresponding conflict probability is about 0.01.
As the time for generating an HAR MaskNet is $2.3\,\text{ms}$, if the tolerable validation time is one minute, the validation can support 1,011 mobiles.

{\em Privacy guarantee:} The neural network masking approach belongs to a broader category of {\em instance encoding}. The formal analysis in a recent study \cite{carlini2021private} has given the theoretical limits of instance encoding in protecting privacy under the notation defined by {\em distinguishing attack}. However, the privacy guarantee of instance encoding under the notations of inversion attack and private attribute extraction is still an open problem. Despite this uncertainty, instance encoding has been increasingly used in recent approaches for resource-constrained devices \cite{osia2017hybrid,wang2018not,li2021deepobfuscator,liu2019privacy,chi2018privacy,malekzadeh2019mobile,hajihassnai2021obscurenet}. This can be due to the practical limitations of other two families of approaches despite their theoretical guarantees \cite{carlini2021private}: cryptographic techniques (including multiparty computation and homomorphic encryption) incurs high computation and communication overheads; differential privacy is typically achieved with high utility losses. Nevertheless, the theoretic limits of neural network masking against inversion attack and private attribute extraction deserve future research.

{\em Distribution shift:} The distribution shift between training dataset and testing dataset often generates impact on the test accuracy in machine learning. For the privacy-preserving propose, the distribution shift can also affect the attack success rate of ExtNet. Thus, the privacy-utility trade-off points may change with the distribution shift. However, the detailed analysis on the impact is slightly out of the scope of this paper. We leave it as an open problem for exploring this issue.
\section{CONCLUSION}
\label{sec:con}

This paper presented PriMask, a cascadable and collusion-resilient data masking approach for mobile devices to use the cloud inference services. In PriMask, the mobile only needs to execute a small-scale neural network called MaskNet to mask the inference data and then sends the result to the cloud. This helps preserve certain private information contained in the inference data. We design a split adversarial learning method to train a neural network used to generate MaskNet for many mobiles. The heterogeneity of MaskNets provides desirable resilience to the potential collusion between any mobile and the cloud. We apply PriMask to three mobile sensing tasks of human activity recognition, urban environment crowdsensing, and driver behavior recognition. The results show PriMask's good generalizability and effectiveness in preserving privacy while maintaining the cloud inference accuracy.


\bibliographystyle{plain}
\balance
\bibliography{refs}

\end{document}